\documentclass[reprint,aps,prb,amsmath,amssymb,citeautoscript]{revtex4-1}
\usepackage[colorlinks=true,citecolor=blue,linkcolor=magenta]{hyperref}
\usepackage[utf8]{inputenc}
\usepackage{graphicx}
\usepackage{lineno}
\usepackage[dvipsnames]{xcolor}
\definecolor{darkred}{HTML}{A80000}

\newcommand{\ie}{\textit{i.e., }}
\begin{document}
\title{Mirror symmetric on-chip frequency circulation of light}

\author{Jason F. Herrmann$^1$}
\email{jfherrm@stanford.edu}
\author{Vahid Ansari$^1$}
\author{Jiahui Wang$^1$}
\author{Jeremy D. Witmer$^1$}
\author{Shanhui Fan$^2$}
\author{Amir H. Safavi-Naeini$^1$}
\affiliation{1. Ginzton Laboratory and Department of Applied Physics, Stanford University, Stanford, CA 94305 USA\\2. Ginzton Laboratory and Department of Electrical Engineering, Stanford University, Stanford, CA 94305 USA}

\date{\today}
\begin{abstract}
    {Integrated circulators and isolators are important for developing on-chip optical technologies, such as laser cavities, communication systems, and quantum information processors. These devices appear to inherently require mirror symmetry breaking to separate backwards from forwards propagation, so existing implementations rely upon magnetic materials, or interactions driven by propagating waves. In contrast to previous work, we demonstrate a mirror symmetric nonreciprocal device. Our device comprises three coupled photonic resonators implemented in thin-film lithium niobate. Applying radio frequency modulation, we drive conversion between the frequency eigenmodes of this system. We measure nearly 40 dB of isolation for approximately 75 mW of RF power near 1550 nm. We simultaneously generate nonreciprocal conversion between all of the eigenmodes in order to demonstrate circulation. Mirror symmetric circulation significantly simplifies the fabrication and operation of nonreciprocal integrated devices. Finally, we consider applications of such on-chip isolators and circulators, such as full-duplex isolation within a single waveguide.}
\end{abstract}    
\maketitle

\begin{figure*}[ht]
    \centering
    \includegraphics[width=0.9\linewidth]{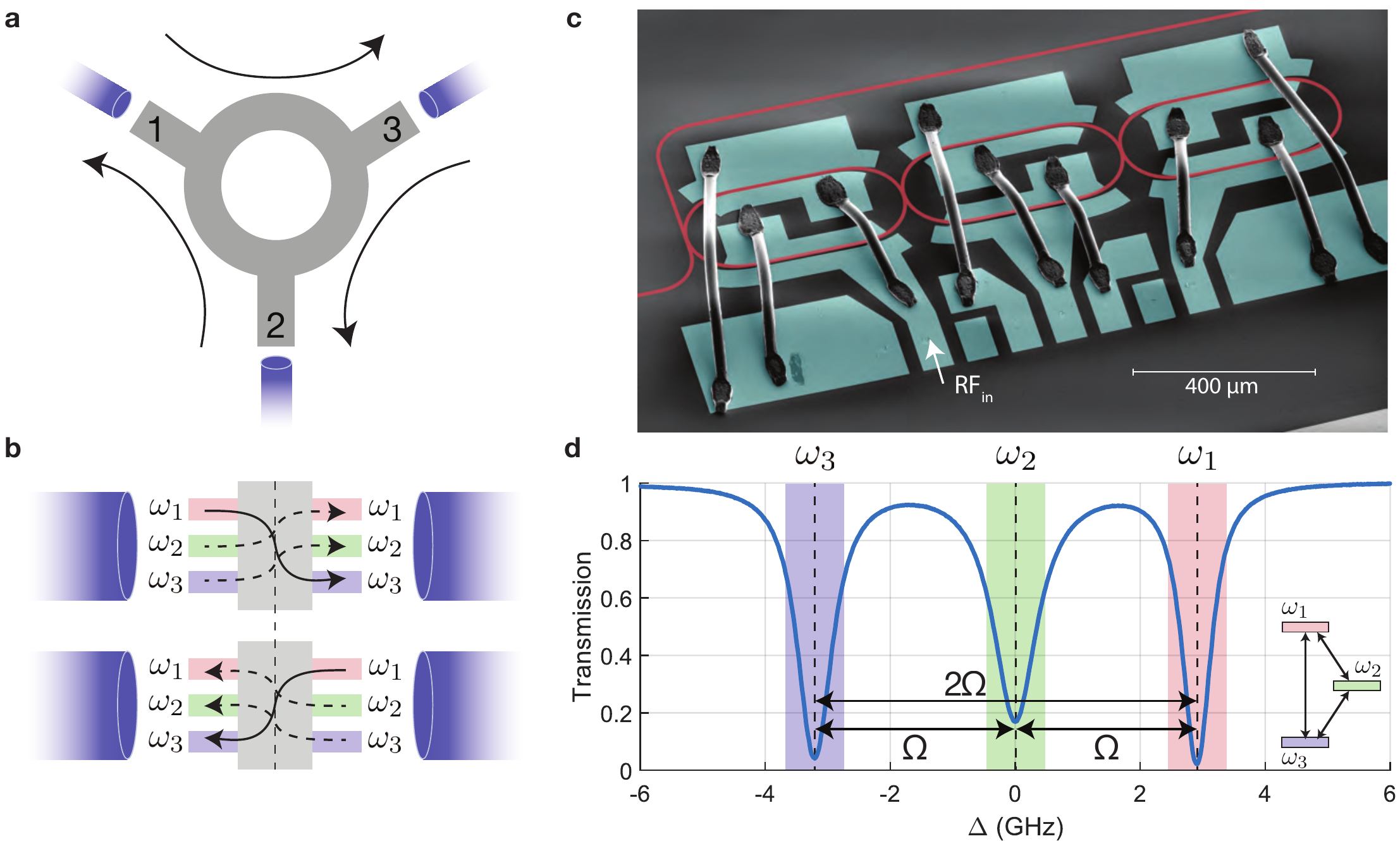}
    \caption{\textbf{Device structure and resonant system.} (a) Schematic of a circulator consisting of three physical waveguide inputs/outputs. (b) Schematic of a circulator in frequency space operating within a single waveguide. The circulator can operate equivalently by inputting light from the left-side waveguide (top) or from the right-side waveguide (bot). The solid arrows emphasize the conversion from $\omega_1$ to $\omega_3$, the dashed arrows indicate alternative frequency conversions, and the vertical dashed line depicts the structural mirror plane of the device. (c) SEM image (false-colored) of an identical device to that used in this experiment. Modulation is applied across both straight-lengths of the first (leftmost) racetrack. DC tuning is applied to the electrodes across the second (middle) racetrack. Lithium niobate (red) and gold electrodes (blue). (d) Optical spectrum of the hybridized super-modes of the device operating at 1543 nm. (d, inset) Schematic of the three-level system formed by the hybridized modes. Frequency conversion between any two modes can occur by two possible pathways; one path is a direct transition, and the other occurs in two steps, mediated by the third level.}
    \label{fig:1_schematic}
\end{figure*}

% AMIR'S VERSION
Circulators are nonreciprocal devices that allow propagating signals to cycle between three channels, \textit{e.g.}, $\text{in}_1\rightarrow\text{out}_3,~ \text{in}_3\rightarrow\text{out}_2,~ \text{in}_2\rightarrow\text{out}_1$ as shown in Fig~\ref{fig:1_schematic}a. A circulator can isolate components from each other, \textit{i.e.}, it can prevent reflections from returning to the source of a signal, simply by terminating one of its three channels through absorption. For example if we terminate port $2$ in Fig. \ref{fig:1_schematic}a, signals will propagate from $\text{in}_1\rightarrow\text{out}_3$, but $\text{in}_3\rightarrow\text{out}_2$ would be terminated, thereby preventing reflections on the third port from propagating back to the first port. There is flexibility in selecting the input and output channels. In many realizations, the channels $\text{in}_k/\text{out}_k$ are the incoming and outgoing waves in the waveguides connected to port $k$. However, other realizations are also possible. In this work, our input channels correspond to the incoming waves at different frequencies in one waveguide (on the left in Fig.~\ref{fig:1_schematic}b), and the output channels correspond to the outgoing waves at different frequencies in the other side of the waveguide (on  the right in Fig.~\ref{fig:1_schematic}b). Unlike more typical three-waveguide configurations, this type of circulator is compatible with \textit{mirror symmetry}~\cite{jiahui2021}: by imposing the same scattering relations between the frequency channels for light traveling from right to left in the device, we obtain a second frequency circulation, related to the first by reflection about the center axis (dashed line in Fig.~\ref{fig:1_schematic}b).

The mirror symmetry of our circulator means that its physical implementation can be simpler than more traditional circulators and isolators, which \emph{require} mirror symmetry breaking. Many of these  demonstrations use magnetic materials~\cite{bi2011chip,tzuang2014non,srinivasan2018magneto,huang2016electrically,yan2020waveguide, sobu2013}, traveling waves~\cite{kittlaus2018non,kittlaus2021non,kim2021chip,shen2016experimental,ruesink2016nonreciprocity,kang2011reconfigurable,kim2015non,dong2015brillouin,hafezi2012optomechanically,tian2021magnetic,sohn2021electrically}, or multiple modulators emitting with different phases~\cite{fang2012photonic, lira2012electrically, doerr2011optical,zu2009,dostart2021optical} to induce a sense of direction and nonreciprocal propagation. As a consequence of mirror symmetry in our device, a single radio frequency (RF) electro-optic modulator (EOM) is sufficient to generate circulation between three optical frequency channels. As we show here, this significantly simplifies hardware implementations of nonreciprocal photonic circuits.

\vspace{15pt}
\noindent\textbf{Results}\medskip\\
We demonstrate a device on a thin film lithium niobate (TFLN) platform that consists of three coupled photonic racetrack resonators. The first resonator is coupled to a bus waveguide, as shown in Fig.~\ref{fig:1_schematic}c. The modes of racetrack resonators $1$ ($2$) and $2$ ($3$) are coupled at a rate $\mu_{12}$ ($\mu_{23}$), while $1$ and $3$ are not directly coupled ($\mu_{13}=0$). Our device uses the $\textrm{TE}_{00}$ guided mode, which on X-cut TFLN has large electro-optic coupling to fields parallel to both the chip surface and the crystal Z axis. Electrodes fabricated across each racetrack enable independent DC bias tuning of the uncoupled or \emph{bare} mode frequencies via the linear electro-optic effect\cite{weisAndgaylord1985}. This is necessary to counteract drift and fabrication disorder.   

We set the desired operating point by using the DC bias to tune all of the bare mode frequencies to be equal, which yields three evenly spaced resonances in the coupled or \emph{dressed} basis with frequencies $\omega_i$. In an ideal system, the spacing is $\Omega = \sqrt{2}\mu$ with $\mu\equiv|\mu_{12}|=|\mu_{23}|$. These three resonances couple to the bus waveguide through their spatial overlaps with the first racetrack and therefore lead to dips in the transmission spectrum, as shown in Fig.~\ref{fig:1_schematic}d. This overlap also means that RF modulation of the racetrack adjacent to the bus waveguide couples all three resonances together, as depicted by the schematic three-level system in Fig.~\ref{fig:1_schematic}d.

We apply an RF modulation of the form:
\begin{equation}
V(t) = A_1\cos(\Omega t + \phi_1) + A_2\cos(2\Omega t + \phi_2).    
\end{equation}
The frequency $\Omega$ corresponds to the microwave modulation frequency. In an ideal  system with equal inter-modal couplings $\mu$, we would set $\Omega$ to $\sqrt{2}\mu$. Fabrication disorder and tuning imprecision means that the $\mu_{ij}$'s are slightly different. For example, $\mu_{12}/2\pi=1.864~\text{GHz}$ and $\mu_{23}/2\pi=1.861~\text{GHz}$, as inferred for the data in Fig.~\ref{fig:1_schematic}d, so we drive at $\Omega/2\pi=2.63~\text{GHz}$, the average of the difference between the measured coupled resonance frequencies $\omega_1-\omega_2$ and $\omega_2-\omega_3$. 
Driving at $\Omega$ scatters light between dressed modes $\omega_1$ and $\omega_2$, whereas the $2\Omega$ drive scatters light between modes $\omega_1$ and $\omega_3$. For any pair of dressed modes, two possible transition pathways exist, as depicted by the schematic inset in Fig. \ref{fig:1_schematic}d. One pathway is a direct transition, whereas the other is a two-step transition through the third mode. By varying the amplitudes and relative phases of the two RF tones, we enhance or suppress different pathways by generating interference. In the ideal disorder-free model, forward isolation (mode $\omega_1\rightarrow\omega_3$) is maximized for the phase condition $2\phi_1 - \phi_2 = \pi/2$.
\begin{figure}[t!]
    \centering
    \includegraphics[width=0.85\columnwidth]{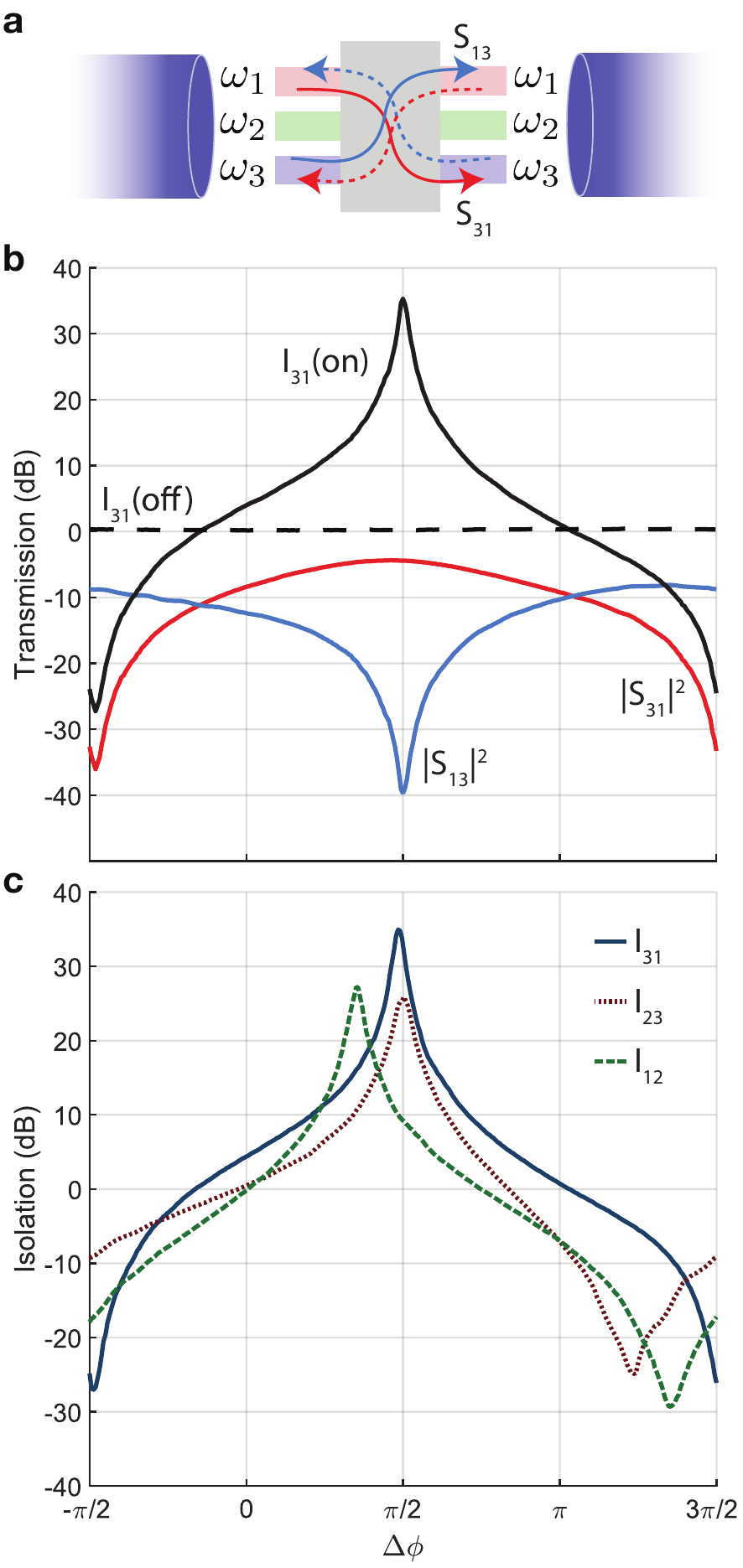}
    \caption{\textbf{Isolation versus microwave phase condition, $\Delta\phi = 2\phi_1 - \phi_2$.} (a) Schematic of frequency-domain circulator. Solid arrows indicate light input from the waveguide on the left (as measured), and dashed arrows indicate the same conversions when light is input from the waveguide on the right. (b) Isolation for $\omega_1\rightarrow\omega_3$ conversion. The line labeled ``off'' corresponds to decreasing the $\Omega$ frequency RF drive ($P_1$) while keeping the $2\Omega$ drive ($P_2$) high, indicating the importance of coherent interference of the two processes for nonreciprocity. $P_1=73.2$ mW, $P_2=59.0$ mW. (c) Mutual isolation for all conversions observed for the fixed power condition in (b). Shifts between the peaks and the ideal phase condition stem from disorder in the mode hybridization.}
    \label{fig:2_MainIso}
\end{figure}\\
\indent We characterize the performance of our circulator by measuring the scattering parameters $S_{ij}$, which quantify how the amplitude $a_j$  of each incoming wave at frequency $\omega_j$ is converted to the amplitude of an outgoing wave $b_i$ at frequency $\omega_i$, propagating in the same direction in the opposite waveguide. This is depicted in Fig. \ref{fig:2_MainIso}a. We define an \emph{isolation parameter} as
\[I_{ij} \equiv |S_{ij}|^2 / |S_{ji}|^2.\] This characterizes the asymmetry between forward and backward frequency conversion for a pair of modes. As shown in Fig.~\ref{fig:2_MainIso}b for channels $1$ and $3$, it is clear that forward scattering from $1\rightarrow3$ occurs efficiently while the backward scattering is strongly suppressed. This leads to $I_{31}$ approaching $40~\text{dB}$ when the driving phase condition is $\pi/2$ and $A_i$ are correctly tuned, as explained below. Similar isolation is observed for other pairs of channels.

\begin{figure}[t!]
    \centering
    \includegraphics[width=\columnwidth]{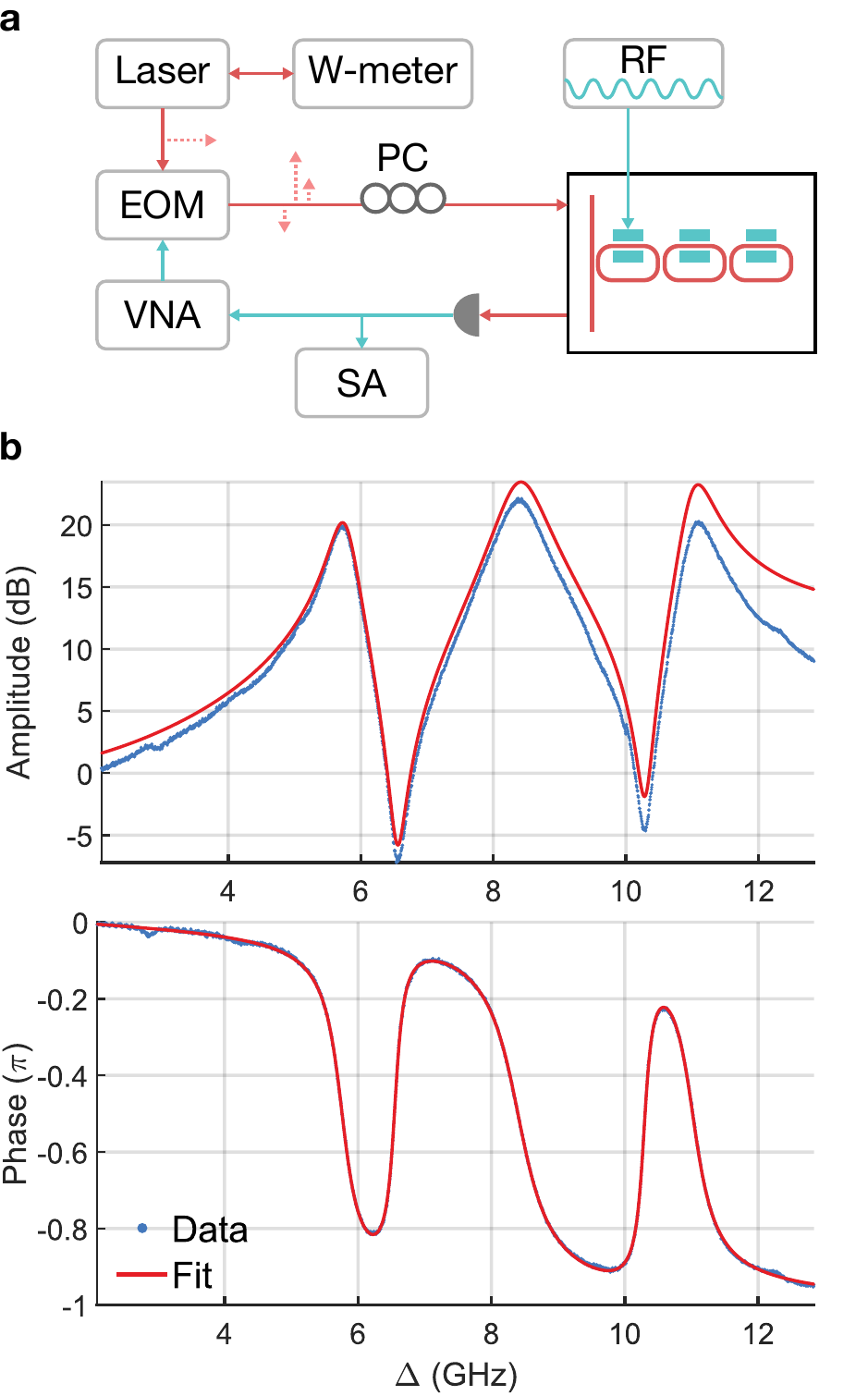}
    \caption{\textbf{Experiment characterization scheme.} (a) Schematic of the characterization setup. (b) Background-normalized amplitude and phase of the VNA spectrum produced by the EOM response filtered through the optical cavity. The EOM is set to an approximately $\pi$ DC phase shift. The fit is made to the phase response and used to predict the amplitude response.
    }
    \label{fig:3_CharacterizationScheme}
\end{figure}

We obtain the scattering and isolation parameters described above by first characterizing the device's linear spectrum, with the RF modulation turned off. We measure the optical transmission parameter $t(\omega)|_\text{RF,off}$ as a function of frequency, which we then use to infer and DC-tune the device's parameters. This transmission amplitude evaluated at frequencies $\omega_j$ corresponds to $S_{jj}|_\text{RF,off}$ since it describes transmission through the device without a change in frequency. We place a laser tone at frequency $\omega_0$,  blue-detuned from all of the resonances, and feed it through a commercial electro-optic modulator (EOM) before sending it to the device. Driving the EOM with a vector network analyzer (VNA), we generate two sidebands at $\omega_0 \pm \Delta$. By sweeping the VNA modulation frequency, we move one sideband across the cavity response, which then beats against the optical feed-through, $\omega_0$, on a subsequent RF photodiode connected to the VNA. Sweeping the VNA frequency allows us to see, nearly in real-time, the hybridized mode structure of the device, as shown in Fig.~\ref{fig:3_CharacterizationScheme}b. Our device exhibits DC bias drift as a result of photorefraction, a common challenge in lithium niobate devices\cite{lisaConverter2020,Jiang2017,Xu2021}. This measurement technique enables us to observe and compensate for these changes as the optical input power and DC bias both affect the spectrum. We fit a complete model of the measurement, including the EOM transmission, cavity response, and phase response measured by the VNA in order to determine all of the device's optical parameters (loss rates, resonant frequencies, etc.). An example of such a fit is depicted in Fig.~\ref{fig:3_CharacterizationScheme}c. We can also extract the sideband transmission $t(\omega)|_\text{RF,off}$  from these spectra. 

We then measure the ``active'' device by turning on the RF drive and characterizing the scattering between frequency channels. Since VNA measurements only find linear scattering parameters, characterizing the scattering between frequencies requires a different measurement scheme than that described above. Keeping the laser at $\omega_0$, we modulate the EOM at frequency $\Delta$ in order to generate and input light at frequency $\omega_j = \omega_0 - \Delta$. We infer all of the optical powers in the different channels by making RF measurements of their beating against the feed-through light at $\omega_0$. This presents a minor complication for inferring $S_{jj}$, as the beat notes generated by both EOM sidebands $\omega_0 \pm \Delta$ interfere at RF frequency $\Delta$. This ambiguity can be resolved by comparing measurements of $S_{jj}$ to the independently measured  scattering parameters (see previous paragraph) when RF modulation to the chip is turned \textit{off}. Note that the other scattering parameters $S_{ij},~i\neq j$ can be inferred without this complication, so the isolation parameters $I_{ij}$ are unaffected. For all scattering parameters, we record the RF power in the beat note corresponding to optical frequency $\omega_i$, while varying the RF phase $\phi_2$. This power is proportional to $|b_i|^2$. We take the ratio of the measured power in $\omega_i$ to that of the unmodulated transmission at the signal frequency $\omega_j$. By factoring out the contribution from the non-resonant input tone at $\omega_0+\Delta$, we obtain the scattering parameters $|S_{ij}|^2 = |b_i/a_j|^2$ (see SI).

The scattering parameter measurements shown in Fig.~\ref{fig:2_MainIso} are taken for optimized values of the RF drive amplitude $A_i$. When the $\Omega$ RF drive is turned down, \textit{i.e.}, $A_1\approx0$, but the $2\Omega$ drive is kept on, we do not observe isolation (dashed line in Fig.~\ref{fig:2_MainIso}b). For tuned values of $A_1$ and $A_2$, we observe simultaneous isolation between the three pairs of dressed frequencies, demonstrating circulator-like behavior, as depicted in Fig.~\ref{fig:2_MainIso}c.

\begin{figure*}[!t]
    \centering
    \includegraphics[width=0.9\linewidth]{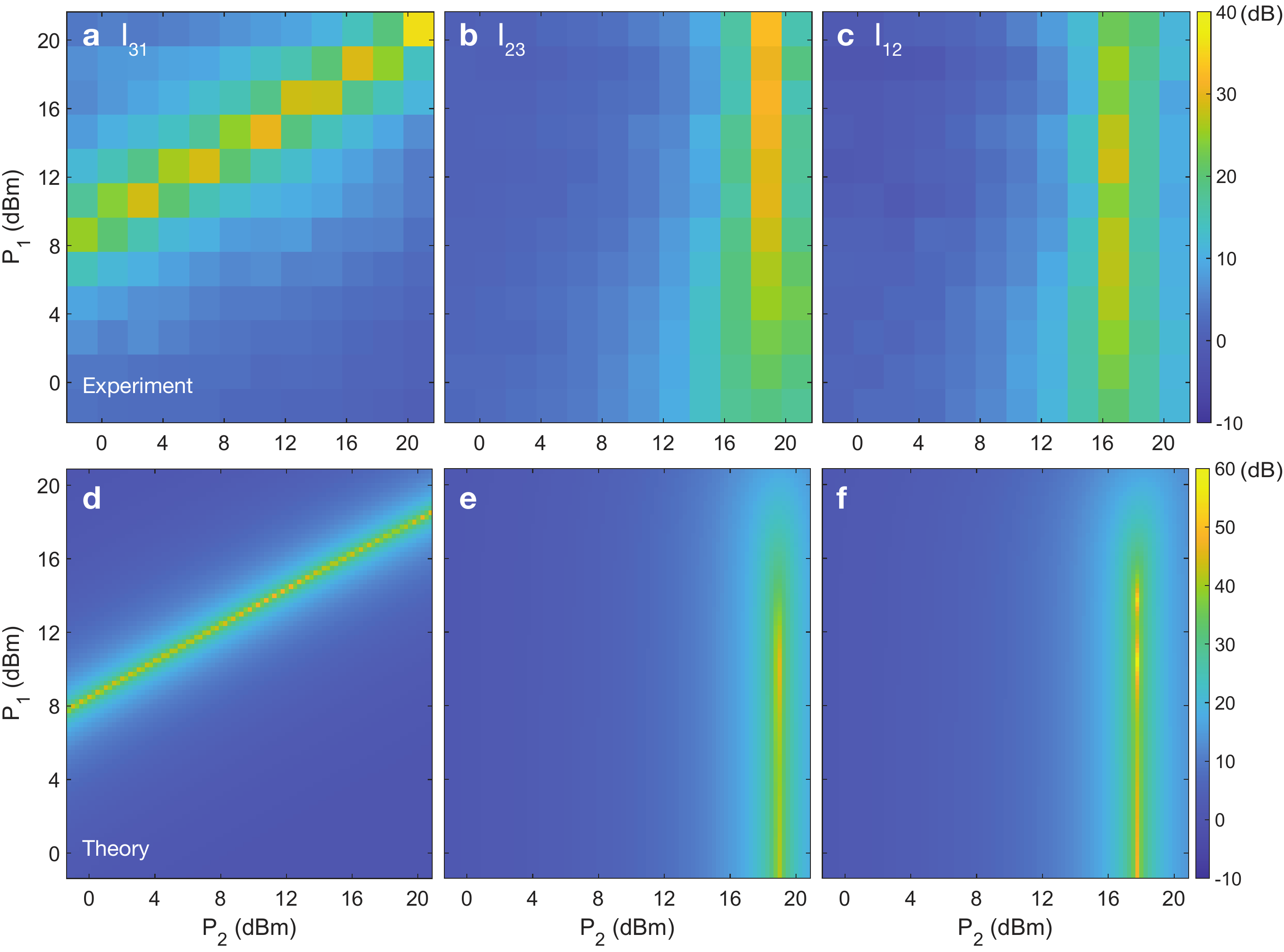}
    \caption{\textbf{Isolation versus microwave power.} We sweep the two RF drive powers, indicated here as $P_1$ ($\Omega$ drive) and $P_2$ ($2\Omega$ drive). For each power combination, the phase of the $2\Omega$ drive ($\phi_2$) is swept across multiple periods. Each pixel corresponds to the maximum observed isolation for each RF power combination for a given isolation parameter. The phase condition varies between pixels of different maps. The top row corresponds to measurements, and the bottom row are theoretical predictions of isolation. We attribute differences between experiment and theory to shifting system parameters over time and as the RF power increases (see SI). Predictions are based on the fit system parameters, which also have some degree of uncertainty.}
    \label{fig:4_HeatMaps}
\end{figure*}

We next characterize the RF power-dependent operation of our device by varying the modulation amplitudes $A_1$ and $A_2$. For each power combination, we sweep the phase condition of the RF sources through multiple periods and extract the peak isolation observed for each transition. We generate maps of the isolation versus RF power, depicted in Fig.~\ref{fig:4_HeatMaps}a-c. Each pixel is normalized by linear scattering parameters, analogously to the data in Fig.~\ref{fig:2_MainIso}. The optical device characteristics used for this normalization are extracted from the VNA trace in Fig.~\ref{fig:3_CharacterizationScheme}, taken prior to varying the RF power. 
%A small arrow in Fig.~\ref{fig:4_HeatMaps}a points to the operating point used for the phase-dependent data shown in Fig.~\ref{fig:2_MainIso}b.

Figures \ref{fig:4_HeatMaps}d-f depict theoretical plots of device performance obtained from coupled mode theory and show good agreement with the measured trends. We attribute differences between theory and experiment, such as the slopes in the $I_{31}$ maps and the locations of peak isolation in $I_{12}$ and $I_{23}$, to mode drifts over the full course of measurement. This drift could emerge from either DC bias drift, which introduces disorder in the mode hybridization, or from high-power RF-induced drift (see SI). We assume this drift does not significantly affect the normalization factor we use for data processing, which is obtained from the initial VNA trace. From the theoretical plots, we infer a high-frequency bare mode modulation rate $g_{\text{EO}}/2\pi\approx 330$ MHz. This rate closely matches the directly measured low-frequency modulation rate $g_{\text{EO,DC}}/2\pi = 328.5$ MHz.

We can understand the trends in the isolation factor RF power dependence by studying the coupled mode theory description of the dynamics\cite{jiahui2021}. The RF modulation scatters photons between the dressed modes at a rate proportional to $\tilde{A}_i = g_{\text{EO}}A_i$, where $g_{\text{EO}}$ is the electro-optic tuning of the bare RF-coupled mode. We observe strong enhancement in $I_{31}$ in Fig.~\ref{fig:4_HeatMaps}a,d along a locus corresponding to $\tilde{A}_1^2 \propto \tilde{A}_2$. This condition is found by noting that for cancellation to occur, the rate at which direct $1\rightarrow 3$ scattering occurs, $\tilde A_2$, needs to match the rate of indirect scattering $1\rightarrow 2 \rightarrow 3$, which is approximately $\tilde A_1^2/\gamma_2$\footnote{By second-order perturbation theory.}. Here, $\gamma_2$ is the linewidth (full-width, half-max -- FWHM) of the second dressed mode. A more detailed analysis\cite{jiahui2021} indicates that maximum $I_{13}$ isolation is given by:
\[\tilde{A}_2=\frac{\tilde{A}_1^2}{2\gamma_2},\]
where $\gamma_2 = (\kappa_1 + \kappa_3)/2$. 

A similar line of reasoning explains the large isolation regions in Fig.~\ref{fig:4_HeatMaps}b,c,e,f. For example, note that a peak $I_{12}$ requires interference between two scattering processes -- a direct process $1\rightarrow 2$, occurring with a rate $\tilde A_1$, and an indirect process $1\rightarrow 3 \rightarrow 2$, occurring at a rate $\tilde A_1 \tilde A_2/\gamma_3$. This interference is maximized when the rates are nearly equal. A more detailed coupled mode analysis confirms that the ideal condition $\tilde{A_2}\tilde{A_1} = 4\gamma_{1,3}\tilde{A_1}$,  where $\gamma_{1,3} = (\kappa_1 + 2\kappa_2 + \kappa_3)/4$, maximizes the isolation parameter. The isolation parameters $I_{12}$ and $I_{23}$ are therefore maximized when:
\[\tilde{A_2} = 4\gamma_{1,3},\]
independently of the direct scattering rate $\tilde A_1$. Inhomogeneity in the mode hybridization causes deviations from this exact condition for $I_{23}$ versus $I_{12}$, \textit{i.e.}, they appear at slightly different power conditions in Fig.~\ref{fig:4_HeatMaps}b,c. %A small arrow in Fig.~\ref{fig:4_HeatMaps} points to the operating point used for the phase-dependent data shown in Fig.~(\ref{fig:2_MainIso}), which simultaneously yields large isolation $I_{31}$, $I_{23}$, and $I_{12}$.%

\vspace{15pt}
\noindent\textbf{Discussion}\medskip\\
Our platform's most significant limit is due to optical loss in the cavities. In the ideal realization the dressed mode linewidths are entirely due to coupling out into the feed waveguide. When the RF drives are off, the absence of intrinsic loss means that the dressed modes are overcoupled, so $|S_{kk}| =1$. As we increase the RF driving power,  the scattering between dressed modes appears as additional loss at the signal frequency. This moves the mode from being over-coupled into a critically coupled condition, when the scattering-induced loss at mode $k$ equals the mode's coupling rate to the waveguide and leads to $|S_{kk}|=0$. However, in the experimentally realized system, there is additional intrinsic loss in all three cavities, and the dressed modes are already close to being critically coupled when the RF drives are off. The RF induced scattering then acts to \emph{increase} transmission $|S_{kk}|$ as it moves the mode farther away from being critically coupled. This means that in contrast to the ideal circulator, the diagonal elements of the scattering matrix are nonzero in our realization. We expect that with improvements in optical $Q$ of on-chip LN devices, this problem can be largely eliminated~\cite{Shams-Ansari:21}. One possible way to circumvent this challenge in the shorter term is to include an additional resonator after the device at frequency $\omega_k$ in order to filter out any feed-through and make $S_{kk}=0$ for a single channel. Another approach that does not require major improvements in the $Q$ is to increase the cavity-waveguide coupling and start within a more over-coupled regime. However, this approach would require larger RF powers and we are already power-limited. The off-chip microwave amplifier saturates at approximately $35$ dBm. Moreover, as power to the device increases above roughly $120$ mW, we observe dressed mode drifts on the order of MHz, which disrupts the resonance condition required for high isolation (see SI).

Another limit of our approach is due to its resonant nature. The isolation bandwidth of our device depends on the dressed mode linewidths, and we measured sustained isolation over more than a few hundred megahertz. This too can be increased, but at the cost of greater RF power as the $\tilde A_{k}$ also scale with bandwidth. One strategy to make the modulation more efficient would be to include modulation into the third racetrack. Alternatively, since we only need narrowband RF modulation, we could use the extremely efficient optomechanical modulation schemes recently demonstrated on LN~\cite{Jiang2020, Shao19, Sarabalis2021} to significantly reduce the needed RF power and realize larger bandwidth. Finally, we note that the bandwidth of the device is potentially larger than the linewidth of a single mode as there are families of modes repeating with the cavity free spectral range, and under low disorder, these would also behave as circulators for signals at different sets of frequencies.

A novel feature of our device is its operation as a \emph{bi-directional} isolator -- a property that emerges from mirror symmetry and frequency-domain operation. For example, two transmitters/receivers can operate simultaneously along a single channel. Node $1$ can transmit a signal down a waveguide at frequency $\omega_1$ to node $2$, who recieves at $\omega_3$. This communication is isolated. Meanwhile, node $2$ can then send a signal backwards along the waveguide at $\omega_1$ to node $1$, who also receives at $\omega_3$. This communication is also isolated. With the standard optical isolator, this behavior would require two optical channels, one for each direction of transmission. 

Overall, we have demonstrated an integrated frequency isolator/circulator on thin-film lithium niobate, an emerging platform for classical and quantum photonics. We measured peak isolation of nearly $40$ dB with $4.4$ dB insertion loss for dual RF drive powers of $P_1 = 73.2$ mW $P_2 = 59.0$ mW. Our device is reconfigurable, enabling isolation over a wide range of powers for different operating frequencies. For example, we also measure $\omega_1\rightarrow\omega_3$ isolation of more than $25$ dB for dual RF powers of $P_1=7.32$ mW and $P_2=0.74$ mW, but with a commensurate increase of insertion loss ($\sim16$ dB). Furthermore, we also measured insertion loss as low as $3$ dB for different power and isolation conditions. Ultimately, the mirror symmetry and frequency-domain operation of our device provide for novel applications as an isolator/circulator and frequency router in photonic circuits.

\vspace{15pt}
\noindent\textbf{Methods}
\medskip\\
\begin{footnotesize}
\noindent \textbf{Fabrication}. The device is fabricated in a two-mask process from 500 nm-thick film of lithium niobate (LN) atop a sapphire (Sa) handle. The components are air-clad, and the electrodes are fabricated as a Ti:Au metal bilayer. The first mask, for photonics fabrication, is defined from hydrogen silsequioxane (HSQ), a negative-tone electron beam resist, patterned with 100 kV electron beam lithography (JEOL JBX-6300FS). We transfer the patterns into the LN with argon ion mill etching (IntlVac ion mill), followed by an acid cleaning procedure. We confirm the device's optical performance prior to fabricating electrodes. The electrodes are defined with a standard photoresist lift-off bilayer. The patterns are written using direct-write lithography (Heidelberg MLA150), and metal is evaporated (Kurt J. Lesker LAB18) prior to solvent-based liftoff. Lastly, we wirebond on-chip to connect electrodes across the optics, thereby defining proper modulation polarity on-chip (West Bond 7476E).\\

\noindent \textbf{Characterization}.
The device is optically pumped from a telecommunications wavelength diode laser (Santec TSL-550). We lock the optical pump blue-detuned from the modes at 1543.6 nm (via a Bristol Wavemeter). The light is passed through a polarization control wheel and in-line fiber polarizer to maximize TE transmission. This light is passed into a commercial electro-optic amplitude modulator (EOM). We drive the EOM from a VNA (R\&S ZNB20) in order to generate sidebands. The pump and sidebands are then passed through another polarization controller, into a variable optical attenuator (VOA), then a power meter before being passed onto the chip. By sweeping the frequency of the VNA output (or by driving a particular frequency) we can sweep the sideband across (or directly drive) the optical dressed modes of the system. The light undergoes modulation on-chip. Output light from the optical waveguide passes through an erbium-doped fiber amplifier (EDFA) and an X-switch, which switches our detector between a photodiode for linear optical characterization, and a fast photo-diode (Optilab PD-40M), which records beat tones between the optical pump, the EOM sideband, and converted sidebands when applying on-chip modulation. The output of the fast photo-diode passes through a bias-tee. One arm passes back to the VNA to record the broad-band response of the cavity, and the other arm passes to a spectrum analyzer (R\&S FSW26) to record converted sideband powers with high precision. This setup enables us to simultaneously observe the full cavity response on the VNA in nearly real-time while adjusting DC bias across the device and manipulating the optical pump, while also observing sideband powers at particular frequencies when driving on-chip modulation.

In order to characterize our device mirror symmetry, we separately insert a second X-switch and polarization control just before the device. We record transmission through the device from both waveguide propagation directions, demonstrating equivalence of scattering matrix elements (see SI).

On-chip modulation is driven from two pulsed signal generators (PSG, Keysight E8257D). One PSG drives the $\Omega$ tone while the other drives the $2\Omega$ tone. We vary the relative phase between these sources. The clocks of the PSGs are locked together, and these are in turn locked to the clock of the FSW26. The two PSG outputs are combined at a power splitter and passed through a high-power microwave amplifier. The output of the amplifier is then passed to a probe, which is contacted to the on-chip electrical pads. Importantly, the rate of direct phase modulation on the PSG is much faster than the global phase drift of the system, enabling us to visualize power modulations at sideband frequencies on the FSW26.\\

\noindent \textbf{Author contribution}. 
J.F.H. fabricated the device. J.F.H. and V.A. led the experimental effort. J.W. developed the device operating theory and characterized theoretical device performance. J.F.H, J.D.W., and J.W. determined physical device designs. J.D.W. assisted in early experimentation. A.H.S.N and S.F. provided experimental and theoretical guidance and support for this experiment.

 \end{footnotesize}

\vspace{15pt}
\noindent\textbf{Acknowledgements}
\medskip
\begin{footnotesize}

\noindent J.F.H. acknowledges support from the National Science Foundation Graduate Research Fellowship Program (Grant No. DGE-1656518). V.A. acknowledges support by the Stanford Q-FARM Bloch Fellowship Program. The authors acknowledge the support of an AFOSR MURI project (FA9550-18-1-0379), the National Science Foundation under award ECCS-1820938, and the Defense Advanced Research Projects Agency (DARPA) LUMOS program. Part of this work was performed at the Stanford Nano Shared Facilities (SNSF), supported by the National Science Foundation under award ECCS-2026822. Work was performed in part in the nano@stanford labs, which are supported by the National Science Foundation as part of the National Nanotechnology Coordinated Infrastructure under award ECCS-1542152. The authors would like to thank Wentao Jiang and Christopher J. Sarabalis for insightful and helpful discussions.
\end{footnotesize}
\bibliography{papers}

\begin{thebibliography}{35}%
\makeatletter
\providecommand \@ifxundefined [1]{%
 \@ifx{#1\undefined}
}%
\providecommand \@ifnum [1]{%
 \ifnum #1\expandafter \@firstoftwo
 \else \expandafter \@secondoftwo
 \fi
}%
\providecommand \@ifx [1]{%
 \ifx #1\expandafter \@firstoftwo
 \else \expandafter \@secondoftwo
 \fi
}%
\providecommand \natexlab [1]{#1}%
\providecommand \enquote  [1]{``#1''}%
\providecommand \bibnamefont  [1]{#1}%
\providecommand \bibfnamefont [1]{#1}%
\providecommand \citenamefont [1]{#1}%
\providecommand \href@noop [0]{\@secondoftwo}%
\providecommand \href [0]{\begingroup \@sanitize@url \@href}%
\providecommand \@href[1]{\@@startlink{#1}\@@href}%
\providecommand \@@href[1]{\endgroup#1\@@endlink}%
\providecommand \@sanitize@url [0]{\catcode `\\12\catcode `\$12\catcode
  `\&12\catcode `\#12\catcode `\^12\catcode `\_12\catcode `\%12\relax}%
\providecommand \@@startlink[1]{}%
\providecommand \@@endlink[0]{}%
\providecommand \url  [0]{\begingroup\@sanitize@url \@url }%
\providecommand \@url [1]{\endgroup\@href {#1}{\urlprefix }}%
\providecommand \urlprefix  [0]{URL }%
\providecommand \Eprint [0]{\href }%
\providecommand \doibase [0]{http://dx.doi.org/}%
\providecommand \selectlanguage [0]{\@gobble}%
\providecommand \bibinfo  [0]{\@secondoftwo}%
\providecommand \bibfield  [0]{\@secondoftwo}%
\providecommand \translation [1]{[#1]}%
\providecommand \BibitemOpen [0]{}%
\providecommand \bibitemStop [0]{}%
\providecommand \bibitemNoStop [0]{.\EOS\space}%
\providecommand \EOS [0]{\spacefactor3000\relax}%
\providecommand \BibitemShut  [1]{\csname bibitem#1\endcsname}%
\let\auto@bib@innerbib\@empty
%</preamble>
\bibitem [{\citenamefont {Wang}\ \emph {et~al.}(2021)\citenamefont {Wang},
  \citenamefont {Herrmann}, \citenamefont {Witmer}, \citenamefont
  {Safavi-Naeini},\ and\ \citenamefont {Fan}}]{jiahui2021}%
  \BibitemOpen
  \bibfield  {author} {\bibinfo {author} {\bibfnamefont {J.}~\bibnamefont
  {Wang}}, \bibinfo {author} {\bibfnamefont {J.~F.}\ \bibnamefont {Herrmann}},
  \bibinfo {author} {\bibfnamefont {J.~D.}\ \bibnamefont {Witmer}}, \bibinfo
  {author} {\bibfnamefont {A.~H.}\ \bibnamefont {Safavi-Naeini}}, \ and\
  \bibinfo {author} {\bibfnamefont {S.}~\bibnamefont {Fan}},\ }\href {\doibase
  10.1103/PhysRevLett.126.193901} {\bibfield  {journal} {\bibinfo  {journal}
  {Phys. Rev. Lett.}\ }\textbf {\bibinfo {volume} {126}},\ \bibinfo {pages}
  {193901} (\bibinfo {year} {2021})}\BibitemShut {NoStop}%
\bibitem [{\citenamefont {Bi}\ \emph {et~al.}(2011)\citenamefont {Bi},
  \citenamefont {Hu}, \citenamefont {Jiang}, \citenamefont {Kim}, \citenamefont
  {Dionne}, \citenamefont {Kimerling},\ and\ \citenamefont
  {Ross}}]{bi2011chip}%
  \BibitemOpen
  \bibfield  {author} {\bibinfo {author} {\bibfnamefont {L.}~\bibnamefont
  {Bi}}, \bibinfo {author} {\bibfnamefont {J.}~\bibnamefont {Hu}}, \bibinfo
  {author} {\bibfnamefont {P.}~\bibnamefont {Jiang}}, \bibinfo {author}
  {\bibfnamefont {D.~H.}\ \bibnamefont {Kim}}, \bibinfo {author} {\bibfnamefont
  {G.~F.}\ \bibnamefont {Dionne}}, \bibinfo {author} {\bibfnamefont {L.~C.}\
  \bibnamefont {Kimerling}}, \ and\ \bibinfo {author} {\bibfnamefont
  {C.}~\bibnamefont {Ross}},\ }\href@noop {} {\bibfield  {journal} {\bibinfo
  {journal} {Nature Photonics}\ }\textbf {\bibinfo {volume} {5}},\ \bibinfo
  {pages} {758} (\bibinfo {year} {2011})}\BibitemShut {NoStop}%
\bibitem [{\citenamefont {Tzuang}\ \emph {et~al.}(2014)\citenamefont {Tzuang},
  \citenamefont {Fang}, \citenamefont {Nussenzveig}, \citenamefont {Fan},\ and\
  \citenamefont {Lipson}}]{tzuang2014non}%
  \BibitemOpen
  \bibfield  {author} {\bibinfo {author} {\bibfnamefont {L.~D.}\ \bibnamefont
  {Tzuang}}, \bibinfo {author} {\bibfnamefont {K.}~\bibnamefont {Fang}},
  \bibinfo {author} {\bibfnamefont {P.}~\bibnamefont {Nussenzveig}}, \bibinfo
  {author} {\bibfnamefont {S.}~\bibnamefont {Fan}}, \ and\ \bibinfo {author}
  {\bibfnamefont {M.}~\bibnamefont {Lipson}},\ }\href@noop {} {\bibfield
  {journal} {\bibinfo  {journal} {Nature photonics}\ }\textbf {\bibinfo
  {volume} {8}},\ \bibinfo {pages} {701} (\bibinfo {year} {2014})}\BibitemShut
  {NoStop}%
\bibitem [{\citenamefont {Srinivasan}\ and\ \citenamefont
  {Stadler}(2018)}]{srinivasan2018magneto}%
  \BibitemOpen
  \bibfield  {author} {\bibinfo {author} {\bibfnamefont {K.}~\bibnamefont
  {Srinivasan}}\ and\ \bibinfo {author} {\bibfnamefont {B.~J.}\ \bibnamefont
  {Stadler}},\ }\href@noop {} {\bibfield  {journal} {\bibinfo  {journal}
  {Optical Materials Express}\ }\textbf {\bibinfo {volume} {8}},\ \bibinfo
  {pages} {3307} (\bibinfo {year} {2018})}\BibitemShut {NoStop}%
\bibitem [{\citenamefont {Huang}\ \emph {et~al.}(2016)\citenamefont {Huang},
  \citenamefont {Pintus}, \citenamefont {Zhang}, \citenamefont {Shoji},
  \citenamefont {Mizumoto},\ and\ \citenamefont
  {Bowers}}]{huang2016electrically}%
  \BibitemOpen
  \bibfield  {author} {\bibinfo {author} {\bibfnamefont {D.}~\bibnamefont
  {Huang}}, \bibinfo {author} {\bibfnamefont {P.}~\bibnamefont {Pintus}},
  \bibinfo {author} {\bibfnamefont {C.}~\bibnamefont {Zhang}}, \bibinfo
  {author} {\bibfnamefont {Y.}~\bibnamefont {Shoji}}, \bibinfo {author}
  {\bibfnamefont {T.}~\bibnamefont {Mizumoto}}, \ and\ \bibinfo {author}
  {\bibfnamefont {J.~E.}\ \bibnamefont {Bowers}},\ }\href@noop {} {\bibfield
  {journal} {\bibinfo  {journal} {IEEE Journal of Selected Topics in Quantum
  Electronics}\ }\textbf {\bibinfo {volume} {22}},\ \bibinfo {pages} {271}
  (\bibinfo {year} {2016})}\BibitemShut {NoStop}%
\bibitem [{\citenamefont {Yan}\ \emph {et~al.}(2020)\citenamefont {Yan},
  \citenamefont {Yang}, \citenamefont {Liu}, \citenamefont {Zhang},
  \citenamefont {Xia}, \citenamefont {Kang}, \citenamefont {Yang},
  \citenamefont {Qin}, \citenamefont {Deng},\ and\ \citenamefont
  {Bi}}]{yan2020waveguide}%
  \BibitemOpen
  \bibfield  {author} {\bibinfo {author} {\bibfnamefont {W.}~\bibnamefont
  {Yan}}, \bibinfo {author} {\bibfnamefont {Y.}~\bibnamefont {Yang}}, \bibinfo
  {author} {\bibfnamefont {S.}~\bibnamefont {Liu}}, \bibinfo {author}
  {\bibfnamefont {Y.}~\bibnamefont {Zhang}}, \bibinfo {author} {\bibfnamefont
  {S.}~\bibnamefont {Xia}}, \bibinfo {author} {\bibfnamefont {T.}~\bibnamefont
  {Kang}}, \bibinfo {author} {\bibfnamefont {W.}~\bibnamefont {Yang}}, \bibinfo
  {author} {\bibfnamefont {J.}~\bibnamefont {Qin}}, \bibinfo {author}
  {\bibfnamefont {L.}~\bibnamefont {Deng}}, \ and\ \bibinfo {author}
  {\bibfnamefont {L.}~\bibnamefont {Bi}},\ }\href@noop {} {\bibfield  {journal}
  {\bibinfo  {journal} {Optica}\ }\textbf {\bibinfo {volume} {7}},\ \bibinfo
  {pages} {1555} (\bibinfo {year} {2020})}\BibitemShut {NoStop}%
\bibitem [{\citenamefont {{Sobu}}\ \emph {et~al.}(2013)\citenamefont {{Sobu}},
  \citenamefont {{Shoji}}, \citenamefont {{Sakurai}},\ and\ \citenamefont
  {{Mizumoto}}}]{sobu2013}%
  \BibitemOpen
  \bibfield  {author} {\bibinfo {author} {\bibfnamefont {Y.}~\bibnamefont
  {{Sobu}}}, \bibinfo {author} {\bibfnamefont {Y.}~\bibnamefont {{Shoji}}},
  \bibinfo {author} {\bibfnamefont {K.}~\bibnamefont {{Sakurai}}}, \ and\
  \bibinfo {author} {\bibfnamefont {T.}~\bibnamefont {{Mizumoto}}},\ }\href
  {\doibase 10.1364/OE.21.015373} {\bibfield  {journal} {\bibinfo  {journal}
  {Opt. Express}\ }\textbf {\bibinfo {volume} {21}},\ \bibinfo {pages} {15373}
  (\bibinfo {year} {2013})}\BibitemShut {NoStop}%
\bibitem [{\citenamefont {Kittlaus}\ \emph {et~al.}(2018)\citenamefont
  {Kittlaus}, \citenamefont {Otterstrom}, \citenamefont {Kharel}, \citenamefont
  {Gertler},\ and\ \citenamefont {Rakich}}]{kittlaus2018non}%
  \BibitemOpen
  \bibfield  {author} {\bibinfo {author} {\bibfnamefont {E.~A.}\ \bibnamefont
  {Kittlaus}}, \bibinfo {author} {\bibfnamefont {N.~T.}\ \bibnamefont
  {Otterstrom}}, \bibinfo {author} {\bibfnamefont {P.}~\bibnamefont {Kharel}},
  \bibinfo {author} {\bibfnamefont {S.}~\bibnamefont {Gertler}}, \ and\
  \bibinfo {author} {\bibfnamefont {P.~T.}\ \bibnamefont {Rakich}},\
  }\href@noop {} {\bibfield  {journal} {\bibinfo  {journal} {Nature Photonics}\
  }\textbf {\bibinfo {volume} {12}},\ \bibinfo {pages} {613} (\bibinfo {year}
  {2018})}\BibitemShut {NoStop}%
\bibitem [{\citenamefont {Kittlaus}\ \emph {et~al.}(2021)\citenamefont
  {Kittlaus}, \citenamefont {Jones}, \citenamefont {Rakich}, \citenamefont
  {Otterstrom}, \citenamefont {Muller},\ and\ \citenamefont
  {Rais-Zadeh}}]{kittlaus2021non}%
  \BibitemOpen
  \bibfield  {author} {\bibinfo {author} {\bibfnamefont {E.~A.}\ \bibnamefont
  {Kittlaus}}, \bibinfo {author} {\bibfnamefont {W.~M.}\ \bibnamefont {Jones}},
  \bibinfo {author} {\bibfnamefont {P.~T.}\ \bibnamefont {Rakich}}, \bibinfo
  {author} {\bibfnamefont {N.~T.}\ \bibnamefont {Otterstrom}}, \bibinfo
  {author} {\bibfnamefont {R.~E.}\ \bibnamefont {Muller}}, \ and\ \bibinfo
  {author} {\bibfnamefont {M.}~\bibnamefont {Rais-Zadeh}},\ }\href@noop {}
  {\bibfield  {journal} {\bibinfo  {journal} {Nature Photonics}\ }\textbf
  {\bibinfo {volume} {15}},\ \bibinfo {pages} {43} (\bibinfo {year}
  {2021})}\BibitemShut {NoStop}%
\bibitem [{\citenamefont {Kim}\ \emph {et~al.}(2021)\citenamefont {Kim},
  \citenamefont {Sohn}, \citenamefont {Peterson},\ and\ \citenamefont
  {Bahl}}]{kim2021chip}%
  \BibitemOpen
  \bibfield  {author} {\bibinfo {author} {\bibfnamefont {S.}~\bibnamefont
  {Kim}}, \bibinfo {author} {\bibfnamefont {D.~B.}\ \bibnamefont {Sohn}},
  \bibinfo {author} {\bibfnamefont {C.~W.}\ \bibnamefont {Peterson}}, \ and\
  \bibinfo {author} {\bibfnamefont {G.}~\bibnamefont {Bahl}},\ }\href@noop {}
  {\bibfield  {journal} {\bibinfo  {journal} {APL Photonics}\ }\textbf
  {\bibinfo {volume} {6}},\ \bibinfo {pages} {011301} (\bibinfo {year}
  {2021})}\BibitemShut {NoStop}%
\bibitem [{\citenamefont {Shen}\ \emph {et~al.}(2016)\citenamefont {Shen},
  \citenamefont {Zhang}, \citenamefont {Chen}, \citenamefont {Zou},
  \citenamefont {Xiao}, \citenamefont {Zou}, \citenamefont {Sun}, \citenamefont
  {Guo},\ and\ \citenamefont {Dong}}]{shen2016experimental}%
  \BibitemOpen
  \bibfield  {author} {\bibinfo {author} {\bibfnamefont {Z.}~\bibnamefont
  {Shen}}, \bibinfo {author} {\bibfnamefont {Y.-L.}\ \bibnamefont {Zhang}},
  \bibinfo {author} {\bibfnamefont {Y.}~\bibnamefont {Chen}}, \bibinfo {author}
  {\bibfnamefont {C.-L.}\ \bibnamefont {Zou}}, \bibinfo {author} {\bibfnamefont
  {Y.-F.}\ \bibnamefont {Xiao}}, \bibinfo {author} {\bibfnamefont {X.-B.}\
  \bibnamefont {Zou}}, \bibinfo {author} {\bibfnamefont {F.-W.}\ \bibnamefont
  {Sun}}, \bibinfo {author} {\bibfnamefont {G.-C.}\ \bibnamefont {Guo}}, \ and\
  \bibinfo {author} {\bibfnamefont {C.-H.}\ \bibnamefont {Dong}},\ }\href@noop
  {} {\bibfield  {journal} {\bibinfo  {journal} {Nature Photonics}\ }\textbf
  {\bibinfo {volume} {10}},\ \bibinfo {pages} {657} (\bibinfo {year}
  {2016})}\BibitemShut {NoStop}%
\bibitem [{\citenamefont {Ruesink}\ \emph {et~al.}(2016)\citenamefont
  {Ruesink}, \citenamefont {Miri}, \citenamefont {Alu},\ and\ \citenamefont
  {Verhagen}}]{ruesink2016nonreciprocity}%
  \BibitemOpen
  \bibfield  {author} {\bibinfo {author} {\bibfnamefont {F.}~\bibnamefont
  {Ruesink}}, \bibinfo {author} {\bibfnamefont {M.-A.}\ \bibnamefont {Miri}},
  \bibinfo {author} {\bibfnamefont {A.}~\bibnamefont {Alu}}, \ and\ \bibinfo
  {author} {\bibfnamefont {E.}~\bibnamefont {Verhagen}},\ }\href@noop {}
  {\bibfield  {journal} {\bibinfo  {journal} {Nature communications}\ }\textbf
  {\bibinfo {volume} {7}},\ \bibinfo {pages} {1} (\bibinfo {year}
  {2016})}\BibitemShut {NoStop}%
\bibitem [{\citenamefont {Kang}\ \emph {et~al.}(2011)\citenamefont {Kang},
  \citenamefont {Butsch},\ and\ \citenamefont
  {Russell}}]{kang2011reconfigurable}%
  \BibitemOpen
  \bibfield  {author} {\bibinfo {author} {\bibfnamefont {M.~S.}\ \bibnamefont
  {Kang}}, \bibinfo {author} {\bibfnamefont {A.}~\bibnamefont {Butsch}}, \ and\
  \bibinfo {author} {\bibfnamefont {P.~S.~J.}\ \bibnamefont {Russell}},\
  }\href@noop {} {\bibfield  {journal} {\bibinfo  {journal} {Nature Photonics}\
  }\textbf {\bibinfo {volume} {5}},\ \bibinfo {pages} {549} (\bibinfo {year}
  {2011})}\BibitemShut {NoStop}%
\bibitem [{\citenamefont {Kim}\ \emph {et~al.}(2015)\citenamefont {Kim},
  \citenamefont {Kuzyk}, \citenamefont {Han}, \citenamefont {Wang},\ and\
  \citenamefont {Bahl}}]{kim2015non}%
  \BibitemOpen
  \bibfield  {author} {\bibinfo {author} {\bibfnamefont {J.}~\bibnamefont
  {Kim}}, \bibinfo {author} {\bibfnamefont {M.~C.}\ \bibnamefont {Kuzyk}},
  \bibinfo {author} {\bibfnamefont {K.}~\bibnamefont {Han}}, \bibinfo {author}
  {\bibfnamefont {H.}~\bibnamefont {Wang}}, \ and\ \bibinfo {author}
  {\bibfnamefont {G.}~\bibnamefont {Bahl}},\ }\href@noop {} {\bibfield
  {journal} {\bibinfo  {journal} {Nature Physics}\ }\textbf {\bibinfo {volume}
  {11}},\ \bibinfo {pages} {275} (\bibinfo {year} {2015})}\BibitemShut
  {NoStop}%
\bibitem [{\citenamefont {Dong}\ \emph {et~al.}(2015)\citenamefont {Dong},
  \citenamefont {Shen}, \citenamefont {Zou}, \citenamefont {Zhang},
  \citenamefont {Fu},\ and\ \citenamefont {Guo}}]{dong2015brillouin}%
  \BibitemOpen
  \bibfield  {author} {\bibinfo {author} {\bibfnamefont {C.-H.}\ \bibnamefont
  {Dong}}, \bibinfo {author} {\bibfnamefont {Z.}~\bibnamefont {Shen}}, \bibinfo
  {author} {\bibfnamefont {C.-L.}\ \bibnamefont {Zou}}, \bibinfo {author}
  {\bibfnamefont {Y.-L.}\ \bibnamefont {Zhang}}, \bibinfo {author}
  {\bibfnamefont {W.}~\bibnamefont {Fu}}, \ and\ \bibinfo {author}
  {\bibfnamefont {G.-C.}\ \bibnamefont {Guo}},\ }\href@noop {} {\bibfield
  {journal} {\bibinfo  {journal} {Nature communications}\ }\textbf {\bibinfo
  {volume} {6}},\ \bibinfo {pages} {1} (\bibinfo {year} {2015})}\BibitemShut
  {NoStop}%
\bibitem [{\citenamefont {Hafezi}\ and\ \citenamefont
  {Rabl}(2012)}]{hafezi2012optomechanically}%
  \BibitemOpen
  \bibfield  {author} {\bibinfo {author} {\bibfnamefont {M.}~\bibnamefont
  {Hafezi}}\ and\ \bibinfo {author} {\bibfnamefont {P.}~\bibnamefont {Rabl}},\
  }\href@noop {} {\bibfield  {journal} {\bibinfo  {journal} {Optics express}\
  }\textbf {\bibinfo {volume} {20}},\ \bibinfo {pages} {7672} (\bibinfo {year}
  {2012})}\BibitemShut {NoStop}%
\bibitem [{\citenamefont {Tian}\ \emph {et~al.}(2021)\citenamefont {Tian},
  \citenamefont {Liu}, \citenamefont {Siddharth}, \citenamefont {Wang},
  \citenamefont {Bl{\'e}sin}, \citenamefont {He}, \citenamefont {Kippenberg},\
  and\ \citenamefont {Bhave}}]{tian2021magnetic}%
  \BibitemOpen
  \bibfield  {author} {\bibinfo {author} {\bibfnamefont {H.}~\bibnamefont
  {Tian}}, \bibinfo {author} {\bibfnamefont {J.}~\bibnamefont {Liu}}, \bibinfo
  {author} {\bibfnamefont {A.}~\bibnamefont {Siddharth}}, \bibinfo {author}
  {\bibfnamefont {R.~N.}\ \bibnamefont {Wang}}, \bibinfo {author}
  {\bibfnamefont {T.}~\bibnamefont {Bl{\'e}sin}}, \bibinfo {author}
  {\bibfnamefont {J.}~\bibnamefont {He}}, \bibinfo {author} {\bibfnamefont
  {T.~J.}\ \bibnamefont {Kippenberg}}, \ and\ \bibinfo {author} {\bibfnamefont
  {S.~A.}\ \bibnamefont {Bhave}},\ }\href@noop {} {\bibfield  {journal}
  {\bibinfo  {journal} {arXiv preprint arXiv:2104.01158}\ } (\bibinfo {year}
  {2021})}\BibitemShut {NoStop}%
\bibitem [{\citenamefont {Sohn}\ \emph {et~al.}(2021)\citenamefont {Sohn},
  \citenamefont {{\"O}rsel},\ and\ \citenamefont
  {Bahl}}]{sohn2021electrically}%
  \BibitemOpen
  \bibfield  {author} {\bibinfo {author} {\bibfnamefont {D.}~\bibnamefont
  {Sohn}}, \bibinfo {author} {\bibfnamefont {O.~E.}\ \bibnamefont {{\"O}rsel}},
  \ and\ \bibinfo {author} {\bibfnamefont {G.}~\bibnamefont {Bahl}},\
  }\href@noop {} {\bibfield  {journal} {\bibinfo  {journal} {arXiv preprint
  arXiv:2104.04803}\ } (\bibinfo {year} {2021})}\BibitemShut {NoStop}%
\bibitem [{\citenamefont {Fang}\ \emph {et~al.}(2012)\citenamefont {Fang},
  \citenamefont {Yu},\ and\ \citenamefont {Fan}}]{fang2012photonic}%
  \BibitemOpen
  \bibfield  {author} {\bibinfo {author} {\bibfnamefont {K.}~\bibnamefont
  {Fang}}, \bibinfo {author} {\bibfnamefont {Z.}~\bibnamefont {Yu}}, \ and\
  \bibinfo {author} {\bibfnamefont {S.}~\bibnamefont {Fan}},\ }\href@noop {}
  {\bibfield  {journal} {\bibinfo  {journal} {Physical review letters}\
  }\textbf {\bibinfo {volume} {108}},\ \bibinfo {pages} {153901} (\bibinfo
  {year} {2012})}\BibitemShut {NoStop}%
\bibitem [{\citenamefont {Lira}\ \emph {et~al.}(2012)\citenamefont {Lira},
  \citenamefont {Yu}, \citenamefont {Fan},\ and\ \citenamefont
  {Lipson}}]{lira2012electrically}%
  \BibitemOpen
  \bibfield  {author} {\bibinfo {author} {\bibfnamefont {H.}~\bibnamefont
  {Lira}}, \bibinfo {author} {\bibfnamefont {Z.}~\bibnamefont {Yu}}, \bibinfo
  {author} {\bibfnamefont {S.}~\bibnamefont {Fan}}, \ and\ \bibinfo {author}
  {\bibfnamefont {M.}~\bibnamefont {Lipson}},\ }\href@noop {} {\bibfield
  {journal} {\bibinfo  {journal} {Physical review letters}\ }\textbf {\bibinfo
  {volume} {109}},\ \bibinfo {pages} {033901} (\bibinfo {year}
  {2012})}\BibitemShut {NoStop}%
\bibitem [{\citenamefont {Doerr}\ \emph {et~al.}(2011)\citenamefont {Doerr},
  \citenamefont {Dupuis},\ and\ \citenamefont {Zhang}}]{doerr2011optical}%
  \BibitemOpen
  \bibfield  {author} {\bibinfo {author} {\bibfnamefont {C.~R.}\ \bibnamefont
  {Doerr}}, \bibinfo {author} {\bibfnamefont {N.}~\bibnamefont {Dupuis}}, \
  and\ \bibinfo {author} {\bibfnamefont {L.}~\bibnamefont {Zhang}},\
  }\href@noop {} {\bibfield  {journal} {\bibinfo  {journal} {Optics letters}\
  }\textbf {\bibinfo {volume} {36}},\ \bibinfo {pages} {4293} (\bibinfo {year}
  {2011})}\BibitemShut {NoStop}%
\bibitem [{\citenamefont {Yu}\ and\ \citenamefont {Fan}(2009)}]{zu2009}%
  \BibitemOpen
  \bibfield  {author} {\bibinfo {author} {\bibfnamefont {Z.}~\bibnamefont
  {Yu}}\ and\ \bibinfo {author} {\bibfnamefont {S.}~\bibnamefont {Fan}},\
  }\href@noop {} {\bibfield  {journal} {\bibinfo  {journal} {Nature Photonics}\
  }\textbf {\bibinfo {volume} {3}},\ \bibinfo {pages} {91} (\bibinfo {year}
  {2009})}\BibitemShut {NoStop}%
\bibitem [{\citenamefont {Dostart}\ \emph {et~al.}(2021)\citenamefont
  {Dostart}, \citenamefont {Gevorgyan}, \citenamefont {Onural},\ and\
  \citenamefont {Popovi{\'c}}}]{dostart2021optical}%
  \BibitemOpen
  \bibfield  {author} {\bibinfo {author} {\bibfnamefont {N.}~\bibnamefont
  {Dostart}}, \bibinfo {author} {\bibfnamefont {H.}~\bibnamefont {Gevorgyan}},
  \bibinfo {author} {\bibfnamefont {D.}~\bibnamefont {Onural}}, \ and\ \bibinfo
  {author} {\bibfnamefont {M.~A.}\ \bibnamefont {Popovi{\'c}}},\ }\href@noop {}
  {\bibfield  {journal} {\bibinfo  {journal} {Optics Letters}\ }\textbf
  {\bibinfo {volume} {46}},\ \bibinfo {pages} {460} (\bibinfo {year}
  {2021})}\BibitemShut {NoStop}%
\bibitem [{\citenamefont {Weis}\ and\ \citenamefont
  {Gaylord}(1985)}]{weisAndgaylord1985}%
  \BibitemOpen
  \bibfield  {author} {\bibinfo {author} {\bibfnamefont {R.~S.}\ \bibnamefont
  {Weis}}\ and\ \bibinfo {author} {\bibfnamefont {T.~K.}\ \bibnamefont
  {Gaylord}},\ }\href {\doibase https://doi.org/10.1007/BF00614817} {\bibfield
  {journal} {\bibinfo  {journal} {Applied Physics A.}\ ,\ \bibinfo {pages}
  {191}} (\bibinfo {year} {1985})}\BibitemShut {NoStop}%
\bibitem [{\citenamefont {McKenna}\ \emph {et~al.}(2020)\citenamefont
  {McKenna}, \citenamefont {Witmer}, \citenamefont {Patel}, \citenamefont
  {Jiang}, \citenamefont {van Laer}, \citenamefont {Arrangoiz-Arriola},
  \citenamefont {Wollack}, \citenamefont {Herrmann},\ and\ \citenamefont
  {Safavi-Naeini}}]{lisaConverter2020}%
  \BibitemOpen
  \bibfield  {author} {\bibinfo {author} {\bibfnamefont {T.~P.}\ \bibnamefont
  {McKenna}}, \bibinfo {author} {\bibfnamefont {J.~D.}\ \bibnamefont {Witmer}},
  \bibinfo {author} {\bibfnamefont {R.~N.}\ \bibnamefont {Patel}}, \bibinfo
  {author} {\bibfnamefont {W.}~\bibnamefont {Jiang}}, \bibinfo {author}
  {\bibfnamefont {R.}~\bibnamefont {van Laer}}, \bibinfo {author}
  {\bibfnamefont {P.}~\bibnamefont {Arrangoiz-Arriola}}, \bibinfo {author}
  {\bibfnamefont {E.~A.}\ \bibnamefont {Wollack}}, \bibinfo {author}
  {\bibfnamefont {J.~F.}\ \bibnamefont {Herrmann}}, \ and\ \bibinfo {author}
  {\bibfnamefont {A.~H.}\ \bibnamefont {Safavi-Naeini}},\ }\href {\doibase
  10.1364/OPTICA.397235} {\bibfield  {journal} {\bibinfo  {journal} {Optica}\
  }\textbf {\bibinfo {volume} {7}},\ \bibinfo {pages} {1737} (\bibinfo {year}
  {2020})}\BibitemShut {NoStop}%
\bibitem [{\citenamefont {Jiang}\ \emph {et~al.}(2017)\citenamefont {Jiang},
  \citenamefont {Luo}, \citenamefont {Liang}, \citenamefont {Chen},
  \citenamefont {Chen},\ and\ \citenamefont {Lin}}]{Jiang2017}%
  \BibitemOpen
  \bibfield  {author} {\bibinfo {author} {\bibfnamefont {H.}~\bibnamefont
  {Jiang}}, \bibinfo {author} {\bibfnamefont {R.}~\bibnamefont {Luo}}, \bibinfo
  {author} {\bibfnamefont {H.}~\bibnamefont {Liang}}, \bibinfo {author}
  {\bibfnamefont {X.}~\bibnamefont {Chen}}, \bibinfo {author} {\bibfnamefont
  {Y.}~\bibnamefont {Chen}}, \ and\ \bibinfo {author} {\bibfnamefont
  {Q.}~\bibnamefont {Lin}},\ }\href {\doibase 10.1364/OL.42.003267} {\bibfield
  {journal} {\bibinfo  {journal} {Opt. Lett.}\ }\textbf {\bibinfo {volume}
  {42}},\ \bibinfo {pages} {3267} (\bibinfo {year} {2017})}\BibitemShut
  {NoStop}%
\bibitem [{\citenamefont {Xu}\ \emph {et~al.}(2021)\citenamefont {Xu},
  \citenamefont {Shen}, \citenamefont {Lu}, \citenamefont {Surya},
  \citenamefont {Sayem},\ and\ \citenamefont {Tang}}]{Xu2021}%
  \BibitemOpen
  \bibfield  {author} {\bibinfo {author} {\bibfnamefont {Y.}~\bibnamefont
  {Xu}}, \bibinfo {author} {\bibfnamefont {M.}~\bibnamefont {Shen}}, \bibinfo
  {author} {\bibfnamefont {J.}~\bibnamefont {Lu}}, \bibinfo {author}
  {\bibfnamefont {J.~B.}\ \bibnamefont {Surya}}, \bibinfo {author}
  {\bibfnamefont {A.~A.}\ \bibnamefont {Sayem}}, \ and\ \bibinfo {author}
  {\bibfnamefont {H.~X.}\ \bibnamefont {Tang}},\ }\href {\doibase
  10.1364/OE.418877} {\bibfield  {journal} {\bibinfo  {journal} {Opt. Express}\
  }\textbf {\bibinfo {volume} {29}},\ \bibinfo {pages} {5497} (\bibinfo {year}
  {2021})}\BibitemShut {NoStop}%
\bibitem [{Note1()}]{Note1}%
  \BibitemOpen
  \bibinfo {note} {By second-order perturbation theory.}\BibitemShut {Stop}%
\bibitem [{\citenamefont {Shams-Ansari}\ \emph {et~al.}(2021)\citenamefont
  {Shams-Ansari}, \citenamefont {Huang}, \citenamefont {He}, \citenamefont
  {Churaev}, \citenamefont {Kharel}, \citenamefont {Tan}, \citenamefont
  {Holzgrafe}, \citenamefont {Cheng}, \citenamefont {Zhu}, \citenamefont {Liu},
  \citenamefont {Desiatov}, \citenamefont {Zhang}, \citenamefont {Kippenberg},\
  and\ \citenamefont {Lon\v{c}ar}}]{Shams-Ansari:21}%
  \BibitemOpen
  \bibfield  {author} {\bibinfo {author} {\bibfnamefont {A.}~\bibnamefont
  {Shams-Ansari}}, \bibinfo {author} {\bibfnamefont {G.}~\bibnamefont {Huang}},
  \bibinfo {author} {\bibfnamefont {L.}~\bibnamefont {He}}, \bibinfo {author}
  {\bibfnamefont {M.}~\bibnamefont {Churaev}}, \bibinfo {author} {\bibfnamefont
  {P.}~\bibnamefont {Kharel}}, \bibinfo {author} {\bibfnamefont
  {Z.}~\bibnamefont {Tan}}, \bibinfo {author} {\bibfnamefont {J.}~\bibnamefont
  {Holzgrafe}}, \bibinfo {author} {\bibfnamefont {R.}~\bibnamefont {Cheng}},
  \bibinfo {author} {\bibfnamefont {D.}~\bibnamefont {Zhu}}, \bibinfo {author}
  {\bibfnamefont {J.}~\bibnamefont {Liu}}, \bibinfo {author} {\bibfnamefont
  {B.}~\bibnamefont {Desiatov}}, \bibinfo {author} {\bibfnamefont
  {M.}~\bibnamefont {Zhang}}, \bibinfo {author} {\bibfnamefont {T.~J.}\
  \bibnamefont {Kippenberg}}, \ and\ \bibinfo {author} {\bibfnamefont
  {M.}~\bibnamefont {Lon\v{c}ar}},\ }in\ \href
  {http://www.osapublishing.org/abstract.cfm?URI=CLEO_SI-2021-STh4J.4} {\emph
  {\bibinfo {booktitle} {Conference on Lasers and Electro-Optics}}}\ (\bibinfo
  {publisher} {Optical Society of America},\ \bibinfo {year} {2021})\ p.\
  \bibinfo {pages} {STh4J.4}\BibitemShut {NoStop}%
\bibitem [{\citenamefont {Jiang}\ \emph
  {et~al.}(2020{\natexlab{a}})\citenamefont {Jiang}, \citenamefont {Sarabalis},
  \citenamefont {Dahmani}, \citenamefont {Patel}, \citenamefont {Mayor},
  \citenamefont {McKenna},\ and\ \citenamefont {van Laer~amd Amir H.
  Safavi-Naeini}}]{Jiang2020}%
  \BibitemOpen
  \bibfield  {author} {\bibinfo {author} {\bibfnamefont {W.}~\bibnamefont
  {Jiang}}, \bibinfo {author} {\bibfnamefont {C.~J.}\ \bibnamefont
  {Sarabalis}}, \bibinfo {author} {\bibfnamefont {Y.~D.}\ \bibnamefont
  {Dahmani}}, \bibinfo {author} {\bibfnamefont {R.~N.}\ \bibnamefont {Patel}},
  \bibinfo {author} {\bibfnamefont {F.~M.}\ \bibnamefont {Mayor}}, \bibinfo
  {author} {\bibfnamefont {T.~P.}\ \bibnamefont {McKenna}}, \ and\ \bibinfo
  {author} {\bibfnamefont {R.}~\bibnamefont {van Laer~amd Amir H.
  Safavi-Naeini}},\ }\href {\doibase
  https://doi.org/10.1038/s41467-020-14863-3} {\bibfield  {journal} {\bibinfo
  {journal} {Nature Communications}\ }\textbf {\bibinfo {volume} {1166}}
  (\bibinfo {year} {2020}{\natexlab{a}}),\
  https://doi.org/10.1038/s41467-020-14863-3}\BibitemShut {NoStop}%
\bibitem [{\citenamefont {Shao}\ \emph {et~al.}(2019)\citenamefont {Shao},
  \citenamefont {Yu}, \citenamefont {Maity}, \citenamefont {Sinclair},
  \citenamefont {Zheng}, \citenamefont {Chia}, \citenamefont {Shams-Ansari},
  \citenamefont {Wang}, \citenamefont {Zhang}, \citenamefont {Lai},\ and\
  \citenamefont {Lon\v{c}ar}}]{Shao19}%
  \BibitemOpen
  \bibfield  {author} {\bibinfo {author} {\bibfnamefont {L.}~\bibnamefont
  {Shao}}, \bibinfo {author} {\bibfnamefont {M.}~\bibnamefont {Yu}}, \bibinfo
  {author} {\bibfnamefont {S.}~\bibnamefont {Maity}}, \bibinfo {author}
  {\bibfnamefont {N.}~\bibnamefont {Sinclair}}, \bibinfo {author}
  {\bibfnamefont {L.}~\bibnamefont {Zheng}}, \bibinfo {author} {\bibfnamefont
  {C.}~\bibnamefont {Chia}}, \bibinfo {author} {\bibfnamefont {A.}~\bibnamefont
  {Shams-Ansari}}, \bibinfo {author} {\bibfnamefont {C.}~\bibnamefont {Wang}},
  \bibinfo {author} {\bibfnamefont {M.}~\bibnamefont {Zhang}}, \bibinfo
  {author} {\bibfnamefont {K.}~\bibnamefont {Lai}}, \ and\ \bibinfo {author}
  {\bibfnamefont {M.}~\bibnamefont {Lon\v{c}ar}},\ }\href {\doibase
  10.1364/OPTICA.6.001498} {\bibfield  {journal} {\bibinfo  {journal} {Optica}\
  }\textbf {\bibinfo {volume} {6}},\ \bibinfo {pages} {1498} (\bibinfo {year}
  {2019})}\BibitemShut {NoStop}%
\bibitem [{\citenamefont {Sarabalis}\ \emph {et~al.}(2020)\citenamefont
  {Sarabalis}, \citenamefont {McKenna}, \citenamefont {Patel}, \citenamefont
  {Van~Laer},\ and\ \citenamefont {Safavi-Naeini}}]{Sarabalis2021}%
  \BibitemOpen
  \bibfield  {author} {\bibinfo {author} {\bibfnamefont {C.~J.}\ \bibnamefont
  {Sarabalis}}, \bibinfo {author} {\bibfnamefont {T.~P.}\ \bibnamefont
  {McKenna}}, \bibinfo {author} {\bibfnamefont {R.~N.}\ \bibnamefont {Patel}},
  \bibinfo {author} {\bibfnamefont {R.}~\bibnamefont {Van~Laer}}, \ and\
  \bibinfo {author} {\bibfnamefont {A.~H.}\ \bibnamefont {Safavi-Naeini}},\
  }\href {\doibase 10.1063/5.0012288} {\bibfield  {journal} {\bibinfo
  {journal} {APL Photonics}\ }\textbf {\bibinfo {volume} {5}},\ \bibinfo
  {pages} {086104} (\bibinfo {year} {2020})},\ \Eprint
  {http://arxiv.org/abs/https://doi.org/10.1063/5.0012288}
  {https://doi.org/10.1063/5.0012288} \BibitemShut {NoStop}%
\bibitem [{\citenamefont {Patel}\ \emph {et~al.}(2018)\citenamefont {Patel},
  \citenamefont {Wang}, \citenamefont {Jiang}, \citenamefont {Sarabalis},
  \citenamefont {Hill},\ and\ \citenamefont {Safavi-Naeini}}]{patel2018}%
  \BibitemOpen
  \bibfield  {author} {\bibinfo {author} {\bibfnamefont {R.~N.}\ \bibnamefont
  {Patel}}, \bibinfo {author} {\bibfnamefont {Z.}~\bibnamefont {Wang}},
  \bibinfo {author} {\bibfnamefont {W.}~\bibnamefont {Jiang}}, \bibinfo
  {author} {\bibfnamefont {C.~J.}\ \bibnamefont {Sarabalis}}, \bibinfo {author}
  {\bibfnamefont {J.~T.}\ \bibnamefont {Hill}}, \ and\ \bibinfo {author}
  {\bibfnamefont {A.~H.}\ \bibnamefont {Safavi-Naeini}},\ }\href {\doibase
  10.1103/PhysRevLett.121.040501} {\bibfield  {journal} {\bibinfo  {journal}
  {Phys. Rev. Lett.}\ }\textbf {\bibinfo {volume} {121}},\ \bibinfo {pages}
  {040501} (\bibinfo {year} {2018})}\BibitemShut {NoStop}%
\bibitem [{\citenamefont {Williamson}\ \emph {et~al.}(2020)\citenamefont
  {Williamson}, \citenamefont {Minkov}, \citenamefont {Dutt}, \citenamefont
  {Wang}, \citenamefont {Song},\ and\ \citenamefont
  {Fan}}]{williamson2020integrated}%
  \BibitemOpen
  \bibfield  {author} {\bibinfo {author} {\bibfnamefont {I.~A.}\ \bibnamefont
  {Williamson}}, \bibinfo {author} {\bibfnamefont {M.}~\bibnamefont {Minkov}},
  \bibinfo {author} {\bibfnamefont {A.}~\bibnamefont {Dutt}}, \bibinfo {author}
  {\bibfnamefont {J.}~\bibnamefont {Wang}}, \bibinfo {author} {\bibfnamefont
  {A.~Y.}\ \bibnamefont {Song}}, \ and\ \bibinfo {author} {\bibfnamefont
  {S.}~\bibnamefont {Fan}},\ }\href@noop {} {\bibfield  {journal} {\bibinfo
  {journal} {Proceedings of the IEEE}\ }\textbf {\bibinfo {volume} {108}},\
  \bibinfo {pages} {1759} (\bibinfo {year} {2020})}\BibitemShut {NoStop}%
\bibitem [{\citenamefont {Jiang}\ \emph
  {et~al.}(2020{\natexlab{b}})\citenamefont {Jiang}, \citenamefont {Sarabalis},
  \citenamefont {Dahmani}, \citenamefont {Patel}, \citenamefont {Mayor},
  \citenamefont {McKenna}, \citenamefont {Van~Laer},\ and\ \citenamefont
  {Safavi-Naeini}}]{wentao2020}%
  \BibitemOpen
  \bibfield  {author} {\bibinfo {author} {\bibfnamefont {W.}~\bibnamefont
  {Jiang}}, \bibinfo {author} {\bibfnamefont {C.~J.}\ \bibnamefont
  {Sarabalis}}, \bibinfo {author} {\bibfnamefont {Y.~D.}\ \bibnamefont
  {Dahmani}}, \bibinfo {author} {\bibfnamefont {R.~N.}\ \bibnamefont {Patel}},
  \bibinfo {author} {\bibfnamefont {F.~M.}\ \bibnamefont {Mayor}}, \bibinfo
  {author} {\bibfnamefont {T.~P.}\ \bibnamefont {McKenna}}, \bibinfo {author}
  {\bibfnamefont {R.}~\bibnamefont {Van~Laer}}, \ and\ \bibinfo {author}
  {\bibfnamefont {A.~H.}\ \bibnamefont {Safavi-Naeini}},\ }\href {\doibase
  https://doi.org/10.1038/s41467-020-14863-3} {\bibfield  {journal} {\bibinfo
  {journal} {Nature Communications}\ }\textbf {\bibinfo {volume} {11}},\
  \bibinfo {pages} {1166} (\bibinfo {year} {2020}{\natexlab{b}})}\BibitemShut
  {NoStop}%
\end{thebibliography}%

\clearpage

% update formatting for same doc
\widetext
\setcounter{page}{1}
\setcounter{equation}{0}
\newcommand{\devdot}[1]{\frac{\delta #1}{\delta t}}
\newcommand{\bsplitter}{\begin{pmatrix} 1 & i\\ i & 1\end{pmatrix}}
\newcommand{\op}[1]{\hat{#1}}
\newcommand{\opDag}[1]{\hat{#1}^{\dagger}}

\renewcommand{\thepage}{S\arabic{page}}
\renewcommand{\thesection}{S\arabic{section}}
\renewcommand{\thetable}{S\arabic{table}}
\renewcommand{\thefigure}{S\arabic{figure}}

\begin{centering}
{\textbf{\large\large{Supplementary Information: Mirror symmetric on-chip frequency circulation of light}}}
\end{centering}
% \author{Jason F. Herrmann$^1$}
% \email{jfherrm@stanford.edu}
% \author{Vahid Ansari$^1$}
% \author{Jiahui Wang$^1$}
% \author{Jeremy D. Witmer$^1$}
% \author{Shanhui Fan$^2$}
% \author{Amir H. Safavi-Naeini$^1$}
% \affiliation{1. Department of Applied Physics, Stanford University, Stanford, CA 94305 USA\\2. Department of Electrical Engineering, Stanford University, Stanford, CA 94305 USA}

% \date{\today}
% \maketitle
\section{Fitting EOM Response}
\subsection{Cavity Response and Coupled Eigenmodes}
\label{section:inputOutput}
We derive the cavity response in the bare-mode basis according to standard coupled mode theory. In the following, $\kappa_i$ describes the total decay rate of bare mode $i$, $\mu_{ij}$ is the coupling between bare modes $i$ and $j$, and $\kappa_e$ is the extrinsic coupling rate of the first ring coupled to the waveguide ($\kappa_1 = \kappa_{1,i} + \kappa_e$). The triple-ring system is described by the Hamiltonian:
\[H = H_0 + H_{int} + H_{WG}\]
\begin{itemize}
    \item $H_0 = \hbar\left(\omega_{0,1}\hat{a}^{\dagger}_1\hat{a}_1+\omega_{0,2}\hat{a}^{\dagger}_2\hat{a}_2+\omega_{0,3}\hat{a}^{\dagger}_3\hat{a}_3\right)$
    \item $H_{int} = \hbar\sum_{i<j} \mu_{ij}\left(\hat{a}^{\dagger}_i\hat{a}_j+\hat{a}^{\dagger}_j\hat{a}_i\right) + H_{WG-int}$
    \item $H_{WG} = -\hbar\int d\omega \gamma(\omega) \hat{b}^{\dagger}(\omega)\hat{b}(\omega)$
\end{itemize}

We take the Heisenberg equations of motion to arrive at the coupled mode theory for the three resonances, $\hat{a}_i$:
\begin{linenomath}
\begin{subequations}
\label{eq:EOMs}
\begin{align}
    \label{eq:a1}
    \dot{\hat{a}}_1(t) &= -\left(i\omega_{0,1} + \frac{\kappa_1}{2}\right)\hat{a}_1-i\mu_{12}\hat{a}_2-\sqrt{\kappa_e}b_{\text{in}}\\
    \label{eq:a2}
    \dot{\hat{a}}_2(t) &= -\left(i\omega_{0,2} + \frac{\kappa_2}{2}\right)\hat{a}_2 - i\mu_{12}\hat{a}_1 - i\mu_{23}\hat{a}_3\\
    \label{eq:a3}
    \dot{\hat{a}}_3(t) &= -\left(i\omega_{0,3} + \frac{\kappa_3}{2}\right)\hat{a}_3 - i\mu_{23}\hat{a}_2\\
    b_{\text{out}} &= b_{\text{in}} + \sqrt{\kappa_e}\hat{a}_1(t)
\end{align}
\end{subequations}
\end{linenomath}
We go into a frame rotating with the laser frequency, $\omega_0$. This shifts our resonances relative to the drive tone, $\omega_{0,i} \to \Delta_i = \omega_{0,i} - \omega_0$. Fourier transforming into the frequency domain and taking $b_{in} = 0$, we obtain the eigenvalue equation \ref{eq:eigenmodes}.\\
\begin{equation}
\label{eq:eigenmodes}
\begin{centering}
\begin{pmatrix}
    -\left(i\Delta_1 + \frac{\kappa_1}{2}\right) & -i\mu_{12} & 0 \\
    -i\mu_{12} & -\left(i\Delta_2 + \frac{\kappa_2}{2}\right) & -i\mu_{23} \\
    0 & -i\mu_{23} & -\left(i\Delta_3 + \frac{\kappa_3}{2}\right)
\end{pmatrix} \Vec{a} = -i\omega\Vec{a}        
\end{centering}
\end{equation}
Here, $\Vec{a}=(\hat{a}_1, \hat{a}_2, \hat{a}_3)^T$. The real part of the eigenvalues corresponds to the dressed mode frequencies $\omega_i$, and the dressed mode loss rates are twice the magnitude of the imaginary part.
\\\\
Taking the Fourier Transform of the equations of motion has the effect of the following further substitutions:
\begin{linenomath}
\begin{gather*}
    \Delta_i \to \Delta_i' = \Delta_i - \Omega\\
  \implies 0 = -\left(i\Delta_1' + \frac{\kappa_1}{2}\right)\hat{a}_1(\Omega)-i\mu_{12}\hat{a}_2(\Omega)-\sqrt{\kappa_e}b_{\text{in}}(\Omega)
\end{gather*}
\end{linenomath}
(and similarly for modes 2,3). We solve the Fourier transformed equations \ref{eq:EOMs} for the linear cavity transmission in the frame rotating with the laser:
\begin{equation}
    \label{eq:r}
        t(\Omega) = \frac{b_{\text{out}}}{b_{\text{in}}} = 1 - \frac{\kappa_e}{i\Delta_1' + \frac{\kappa_1}{2}+\frac{\mu_{12}^2}{i\Delta_2' + \frac{\kappa_2}{2} + \frac{\mu_{23}^2}{i\Delta_3' + \frac{\kappa_3}{2}}}}
\end{equation}
\subsection{EOM Response}
We calculate the photodiode response to light modulated by an EOM traveling through our device\cite{patel2018}. An intensity modulator can be described as a $50/50$ beam-splitter, followed by a variable phase shift on one path, followed by a second $50/50$ beam-splitter. This is given by the following.
\begin{equation}
    (\alpha_\text{out},\cdot)^T = \frac{1}{2}\bsplitter
    \begin{pmatrix}
        b & 0\\ 0 & ae^{i\phi}
    \end{pmatrix}\bsplitter(\alpha_\text{in}, 0)^T
\end{equation}
In the above, $(\alpha_\text{in}, 0)^T$ is the laser input in the rotating frame of the laser. Furthermore, $b,a$ describe asymmetric loss between the arms of the integrated modulator. The phase accumulated in the EOM is given by a DC phase shift, $\theta$, plus harmonic modulation:
\begin{equation}
    \phi = \theta + \beta\cos(\Omega t+\phi')
\end{equation}
$\beta$ is small, so we can neglect higher-order sidebands. We solve for the time-dependent transmission through the EOM:
\begin{equation}
     \alpha_\text{out} = \frac{\alpha_\text{in}}{2}\left[(b-ae^{i\theta})-\frac{i\beta}{2}e^{i\theta}e^{i\phi'}e^{i\Omega t}-\frac{i\beta}{2}e^{i\theta}e^{-i\phi'}e^{-i\Omega t}\right]
\end{equation}
In our experiment, the laser is blue-detuned from the cavity, so the $-|\Omega|$ sideband sweeps across a mode as we sweep the modulation frequency $\Omega$. The carrier and both sidebands are then filtered by the linear cavity response in eq. \ref{eq:r} to yield the time-dependent transmission through our device. This leads to an output field amplitude:
\begin{equation}
\label{eq:inputPower}
    \alpha_\text{out} = \frac{\alpha_\text{in}}{2}\left[(a-e^{i\theta})t(0)-\frac{i\beta}{2}e^{i\theta}e^{i\phi'}e^{i\Omega t}t(-\Omega)-\frac{i\beta}{2}e^{i\theta}e^{-i\phi'}e^{-i\Omega t}t(+\Omega)\right]
\end{equation}
This transmitted field is recorded on a photodiode and routed to a VNA. We AC-couple the output to neglect DC offsets: \begin{equation}\label{eq:intensityOut}
\begin{split}
    |\alpha_\text{out}|^2 &= \frac{i|\alpha_\text{in}|^2\beta}{8}e^{i\phi'}e^{i\Omega t}[(a-e^{i\theta})e^{-i\theta}t(0)t^*(+\Omega)
    - (a-e^{-i\theta})e^{i\theta}t^*(0)t(-\Omega)] + c.c.\\
    &= Z(\Omega)e^{i\Omega t} + Z^*(\Omega)e^{-i\Omega t}
\end{split}
\end{equation}
The voltage from the photodiode is given by $V(t) \sim \mathcal{R}\{Z(\Omega)e^{i\Omega t}\}$, and the VNA trace is proportional to the amplitude of the comlex number $Z(\Omega)$.

We normalize by the far-detuned response ($t(\omega)=t(0)\approx1, \Delta\gg1$), which allows us to cancel pre-factors corresponding to the carrier tone amplitude, RF modulation amplitude, global phases, and frequency-dependent cable and detector losses. We can fit the normalized phase response of the VNA ($\text{arg}\{Z(\Omega)\}$) to determine system parameters.

\begin{center}
\begin{table}[!t]
\label{tb:fitparams}
\caption{System parameters, measured and fit to VNA phase trace}
\begin{tabular}{c|c|r|r}
\hline
     Parameter & Description & Average & 95\% Confidence Interval \\ 
     \hline\hline
     $\omega_L$ & Wavelength of laser drive & $1543.605$ nm  & -\\
     $\kappa_e/2\pi$ & Extrinsic loss rate of the first resonator & $847$ MHz & $[711, 1002]$ MHz \\
     $\kappa_{i}/2\pi$ & Intrinsic loss rate of the first resonator & $404$ MHz & $[249, 540]$ MHz \\
     $\kappa_{1}/2\pi$ & First resonator total loss rate ($\kappa_1 = \kappa_e + \kappa_i$) & $1.251$ GHz & $[1.247, 1.254]$ GHz \\
     $\kappa_2/2\pi$ & Second resonator total loss rate &$198$ MHz & $[196, 200]$ MHz\\
     $\kappa_3/2\pi$ & Third resonator total loss rate & $225$ MHz & $[222, 228]$ MHz \\
     $\mu_{12}/2\pi$ & Coupling rate between first and second resonators & $1.864$ GHz & $[1.863, 1.865]$ GHz\\
     $\mu_{23}/2\pi$ & Coupling rate between second and third resonators &$1.861$ GHz & $[1.861, 1.862]$ GHz\\
     $\Delta_1/2\pi$ & Detuning of first bare mode from laser drive & $-8.616$ GHz & $[-8.621, -8.611]$ GHz\\
     $\Delta_2/2\pi$ & Detuning of second bare mode from laser drive & $-8.384$ GHz & $[-8.386, -8.383]$ GHz\\
     $\Delta_3/2\pi$ & Detuning of third bare mode from laser drive & $-8.175$ GHz & $[-8.180, -8.170]$ GHz\\
     $\omega_3/2\pi$ & Detuning of third dressed mode from laser drive & $-11.016$ GHz & $[-11.016, -11.015]$ GHz\\
     $\omega_2/2\pi$ & Detuning of second dressed mode from laser drive & $-8.400$ GHz & $[-8.401, -8.399]$ GHz\\
     $\omega_1/2\pi$ & Detuning of first dressed mode from laser drive & $-5.761$ GHz & $[-5.761, -5.760]$ GHz\\
     $\Omega/2\pi$ &  RF drive frequency & $2.63$ GHz & -\\
     $P_{1}$ & RF power used in below detuning study & 73.2 mW & -\\
     $P_{2}$ & RF power used in below detuning study & 59.0 mW & -\\
     $g_{\text{EO,RF}}/2\pi$ & High frequency bare mode modulation rate (inferred) & 330 MHz/V & -\\
     $g_{\text{EO,DC}}/2\pi$ & Low frequency bare mode modulation rate (measured) & 328.5 MHz/V & -\\
\end{tabular}
\end{table}
\end{center}
\subsection{Bootstrapping: System Parameters}
\label{section:bootstrapping}
Our model has many free parameters, so our fit quality is sensitive to an initial guess. We account for this by applying a modified bootstrap algorithm. We repeatedly fit subsets of the VNA trace using random parameter guesses within physical bounds. We obtain distributions of the fit parameters, summarized in table \ref{tb:fitparams}. After fitting $\Delta_1$, $\Delta_2$, $\Delta_3$, we can solve eq. \ref{eq:eigenmodes} for the dressed mode detunings, $\omega_i$, from the carrier tone. We generate input signals at dressed mode frequencies by driving the EOM at $|\omega_i|$. When the EOM DC bias phase $\theta \approx \pi$, the VNA trace appears symmetric, and we can approximate the super-mode locations by extracting the peaks in the VNA trace.
% \begin{table}[]
% \label{tb:eigenmodes}
% \centering
%     \caption{Approximate resonant mode locations obtained by peaks of VNA trace.}
%     \begin{tabular}{r|r}
%     \hline
%      Approximate Frequency & GHz \\ 
%      \hline\hline
%      $W_3'/2\pi$ & -11.065 \\
%      $W_2'/2\pi$ & -8.363 \\
%      $W_1'/2\pi$ & -5.672 \\
% \end{tabular}
% \end{table}
In order to improve the symmetry of the mode transitions in the conversion experiment, we apply RF modulation at $\Omega = (\omega_1-\omega_3)/2$, and we position the $\omega_2$ input at $(\omega_1+\Omega)/2\pi = 8.391$ GHz.
% However, if the modes are not symmetric, then optically pumping at the same frequency as the RF drive tone will yield more symmetric/equal inter-band transitions, whereas pumping at the calculated eigenfrequency would introduce additional laser-microwave detuning that would distort the converted sideband shapes and reduce isolation.
\section{Scattering Matrix Formalism}
\subsection{Normalization and Scattering Parameters}
In order to establish non-reciprocity, we must demonstrate that the scattering matrix describing input and output of the device is asymmetric\cite{williamson2020integrated}. We first consider the forward (left-to-right) propagating modes. In the main text, we describe our measurement procedure to obtain scattering matrix elements. We measure the linear transmission amplitude, corresponding to un-modulated signal light, $a_j$. The power recorded through the photodiode onto an FSW, $\tilde{a}_j$ includes contributions from both the resonant EOM sideband (\ie the signal input at frequency $\omega_j = \omega_0 - \Delta$) and the non-resonant sideband ($\omega_0+\Delta$). The measured power is related to the input signal amplitude by a normalization factor: $a_j = \eta_j\tilde{a}_j$. Turning on RF modulation, we measure the converted sideband amplitudes $b_i$. Scattering is given by the expression:
\begin{equation}
        \vec{b}_i =\mathbf{S_{ij}}\vec{a_j} =
    \begin{pmatrix}
        S_{11} & S_{12} & S_{13} \\
        S_{21} & S_{22} & S_{23} \\
        S_{31} & S_{32} & S_{33}
    \end{pmatrix}
    \begin{pmatrix}
        \eta_1\tilde{a}_1\\
        \eta_2\tilde{a}_2\\
        \eta_3\tilde{a}_3
    \end{pmatrix}
\end{equation}
% The vector $\vec{b}$ describes the output amplitudes in the three super-modes under on-chip modulation, whereas $\vec{b}_{\text{off}}$ describes the transmitted signal power at the three super-modes under \textit{no modulation}. The un-modulated transmission under no modulation includes contributions from both the red and blue sidebands of the EOM positioned at the input to the device. We measure the modulated sideband power, $|b_j|^2$ directly on an FSW. 
The power in a given sideband is proportional to the integration of the autocorrelation function $S_{VV}$, which is recorded on the FSW. This integration is proportional to $|Z(\Omega)|^2$ from \ref{eq:intensityOut}. We therefore measure sideband power as a function of the transmitted signal power:
\begin{equation}
\begin{split}
    \label{eq:SjiPrime}
    |S_{ij}|^2 &= \frac{|b_i|^2}{|Z(|\omega_j|)\eta(\omega_j)|^2}\\
    &\propto \frac{|b_i|^2}{|(a-e^{i\theta})e^{-i\theta}t(0)t^*(|\omega_j|) - (a-e^{-i\theta})e^{i\theta}t^*(0)t(-|\omega_j|)|^2|\eta_j|^2}
\end{split}
\end{equation}
\noindent We identify $\eta_j$ such that we factor out the non-resonant sideband contribution in \ref{eq:SjiPrime}. 
\begin{equation}
    \begin{split}
        \label{eq:eta}
        |\eta_j|^2 &= \frac{|(a-e^{-i\theta})e^{i\theta}t^*(0)|^2}{|(a-e^{i\theta})e^{-i\theta}t(0)t^*(|\omega_j|) - (a-e^{-i\theta})e^{i\theta}t^*(0)t(-|\omega_j|)|^2}
    \end{split}
\end{equation}

\noindent All of the free parameters in equations \ref{eq:SjiPrime} and \ref{eq:eta} are obtained via the fits demonstrated in section \ref{section:bootstrapping}. The diagonal elements $S_{jj}$ also contain a non-resonant contribution in the transmitted power measurement, so they require a slightly modified normalization factor.

For a fixed phase condition, $\Delta\phi=2\phi_1 - \phi_2$, and modulation amplitudes, $P_1 = 73.2$ mW and $P_2 = 59.0$ mW, the scattering matrix $\mathbf{S}$ takes on an asymmetric form demonstrating circulation. The diagonal elements correspond to feedthrough at the original signal frequency as described in the main text. Here we present the scattering matrices for both forward ($1\to3\to2\to1$) and reverse ($1\to2\to3\to1$) circulation for left-to-right-propagating signals. 

\begin{equation}
    \label{eq:sForward}
    \Delta\phi = +\frac{\pi}{2}\implies |S_+|^2 =\begin{pmatrix}
        0.21 & 0.21 & 0.00 \\
        0.02 & 0.19 & 0.26 \\
        0.36 & 0.00 & 0.29
    \end{pmatrix}
\end{equation}
\begin{equation}
    \Delta\phi = -\frac{\pi}{2}\implies |S_-|^2=\begin{pmatrix}
        0.19 & 0.01 & 0.13 \\
        0.35 & 0.25 & 0.05 \\
        0.00 & 0.30 & 0.38
    \end{pmatrix}
\end{equation}
The scattering matrices are asymmetric, thereby demonstrating non-reciprocal frequency conversion and amplitude transmission between the frequency ports. Diagonal elements correspond to feed-through at the original signal frequencies, which can be reduced using techniques discussed in the main text.

% If we include the right-to-left propagating frequency modes (\ie inputs from the right-hand side of the waveguide), our system expands to six ports, and the full system is described as follows:
% \begin{equation}
%     \begin{pmatrix}
%         \vec{b}_1 \\
%         \vec{b}_2
%     \end{pmatrix}=\mathbf{S}
%     \begin{pmatrix}
%         \vec{a}_1\\
%         \vec{a}_2
%     \end{pmatrix}=
%     \begin{pmatrix}
%         R_f & T_b \\
%         T_f & R_b
%     \end{pmatrix}
%     \begin{pmatrix}
%         \vec{a}_1\\
%         \vec{a}_2
%     \end{pmatrix}
% \end{equation}
% In this case, $\vec{a}_i = (a_{i,1},a_{i,2},a_{i,3})^T$ and $\vec{b}_i=(b_{i,1},b_{i,2},b_{i,3})^T$, describe inputs and outputs for the three super-mode frequencies at physical channels (i.e., propagation directions) $i\in[1,2]$.
% \\\\
% $S$ is a six-by-six matrix composed of 3x3 sub-matrices, $R$ and $T$, which relate reflection and transmission of the three frequencies between the two physical ports. Subscripts $b$ and $f$ describe backward- (left) and forward- (right) propagating modes in the waveguide. In a mirror-symmetric device, $T_f = T_b$. Therefore, by characterizing conversion for a single channel input, we characterize conversion from either waveguide directional input.

\subsection{Forward and Reverse Isolation}
\begin{figure}[hb]
    \centering
    \includegraphics[width=0.6\linewidth]{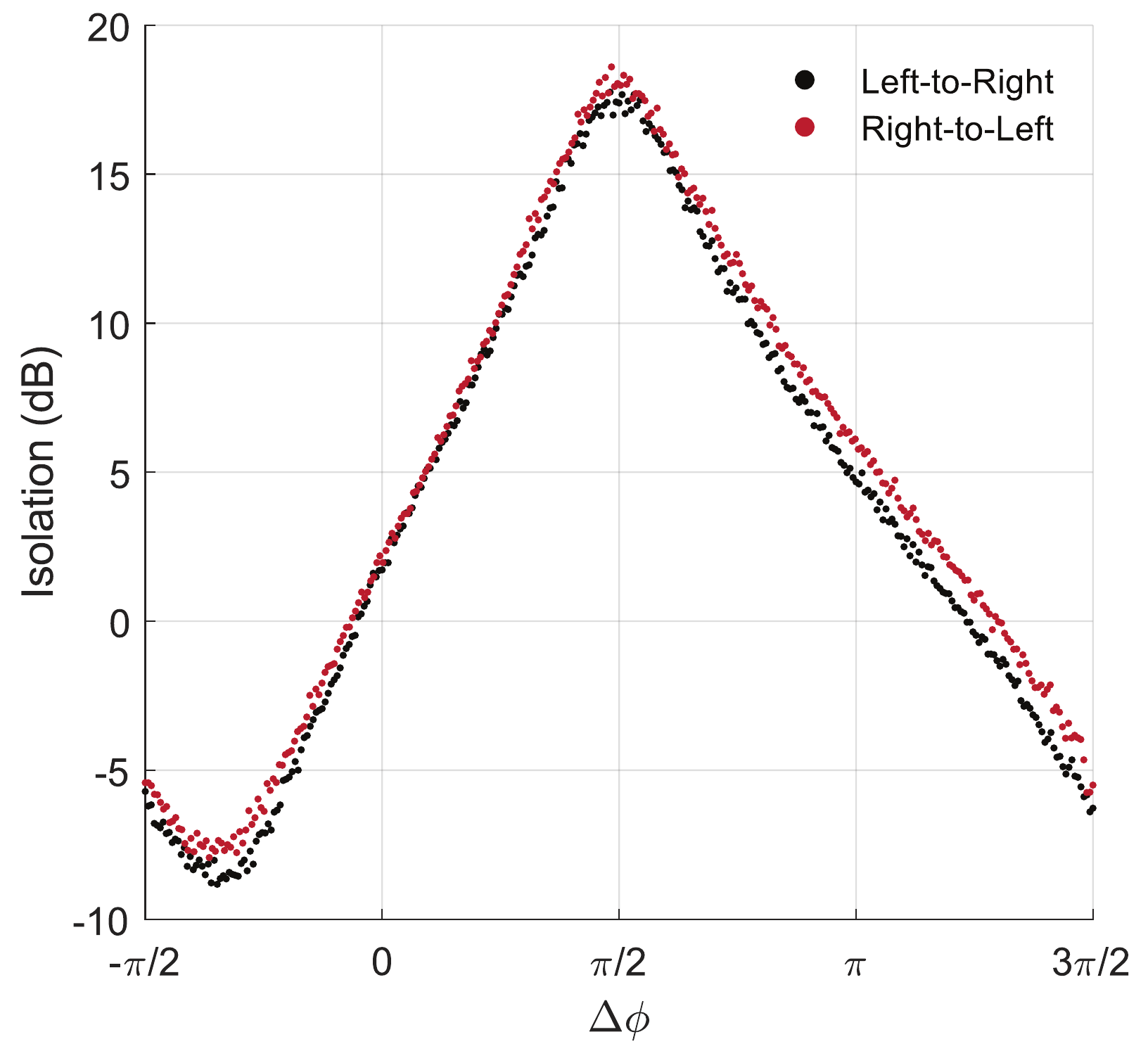}
    \caption{\textbf{Mirror symmetric isolation, $I_{31}$.} Inputs propagating from left-to-right versus right-to-left in the bus waveguide yield equivalent sideband conversion and isolation. Fluctuations in the transmission and conversion are due to slight differences in the polarization of input light from either direction.}
    \label{fig:MirrorSymmetry}
\end{figure}

We confirm mirror symmetry in our device by measuring and comparing isolation for light propagating from left-to-right versus right-to-left in the feed waveguide. We insert an ``x-switch'' and additional polarization controller before the device in order to swap the waveguide input direction. We match polarization of the input light between the two paths and measure $I_{31}$, depicted in Fig.~\ref{fig:MirrorSymmetry}. Isolation is equivalent for light propagating from either direction in the bus waveguide. This validates our single-port characterization scheme. We attribute slight discrepancies between the propagation directions to fluctuations in the polarization on each path. This measurement was taken for RF drive powers $P_1=74.7$ mW and $P_2=65.6$ mW and for RF drive frequency $\Omega=2.64$ GHz.

\section{Pump Depletion}
We characterize the effects of RF modulation on mode drift by applying RF modulation at $\Omega/2\pi = (\omega_1-\omega_3)/2\pi = 5.26$ GHz and recording a series of spectra on the VNA. After collecting several traces, we turn off modulation and record another series of traces. By recording the time stamp of each scan, we measure mode drifts in near real-time. 
% We propose using such a scheme in future studies to characterize photorefractive or pyroelectric effects in thin-film lithium niobate photonics.

\begin{figure*}[!t]
    \centering
    \includegraphics[width=0.8\linewidth]{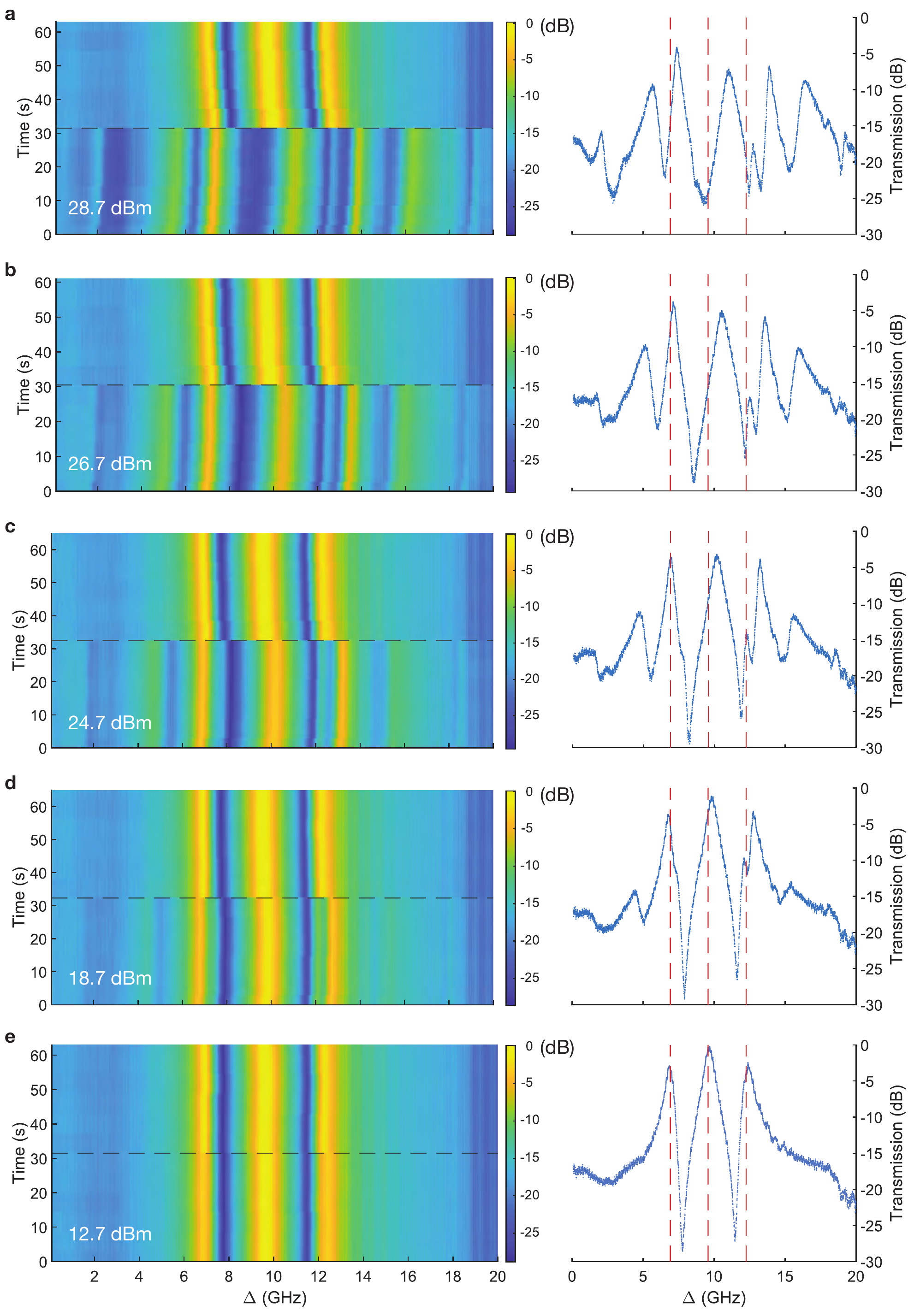}
    \caption{\textbf{Mode drift and pump depletion under RF modulation.} Repeated VNA spectra taken over time for the RF drive powers indicated on each heatmap. The dashed black lines in each heatmap indicate the time at which modulation is turned off. Plots to the right correspond to the VNA trace taken just before modulation is turned off (i.e., the horizonatal slice of the heatmap at the dashed black line). Red dashed lines in the traces correspond to the initial peak locations prior to turning on any RF modulation.}
    \label{fig:S1_PumpDepletion}
\end{figure*}

Figure \ref{fig:S1_PumpDepletion} presents scans taken at various RF powers. The black dashed line indicates the first scan for which modulation is turned off in each set. The individual traces correspond to the last scan under RF modulation. We normalize these traces to the maximum amplitude of the modes before \textit{any} modulation is applied (\ie before taking data in Fig.~\ref{fig:S1_PumpDepletion}e). Operating our EOM at the DC $\pi$ phase point, we approximate peaks in the spectra as mode locations. We track the center mode's location and observe drift on the order of half of a linewidth under high-power modulation. After turning off RF modulation, the modes relax to their initial locations. There is an initial offset drift in the mode location at each RF power, which we believe occurs on a timescale shorter than the speed at which we extract data from the VNA.

This measurement scheme is useful for in situ mode characterization, as the mode behavior can be observed nearly independently of such effects as the optical power or the laser sweep rate (which changes intra-cavity optical power). The VNA/EOM sweeps are rapid and low-power, yielding an almost constant optical energy into the modes, and visualizing what appears to be steady-state mode behavior. This steady-state changes for different laser pump amplitudes, DC bias voltage, or RF power, but is independent of the VNA sweep rate or power applied to the EOM. We believe mode drift and distortion are functions of pump depletion and on-chip heating at high RF modulation powers.
\begin{figure*}[!h]
    \centering
    \includegraphics[width=0.9\linewidth]{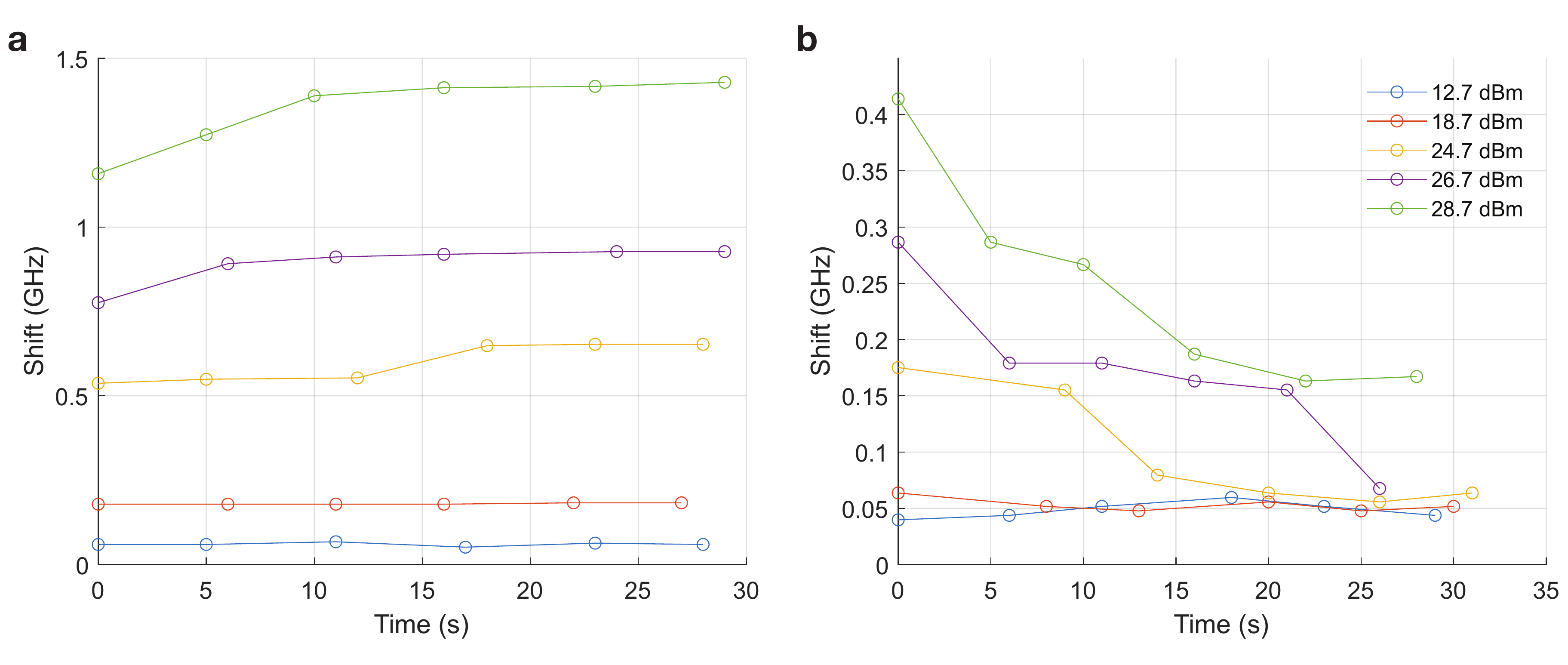}
    \caption{\textbf{Mode drift as a function of time.} (a) Drift under RF modulation. (b) Relaxation when RF modulation is turned off. In both cases, the shift corresponds to drift of the central mode peak relative to its intial frequency prior to any modulation.}
    \label{fig:S1_PumpDepletion_Summary}
\end{figure*}
\clearpage
\section*{S4 Detuning Effects}
\subsection{Microwave Detuning}
It is inefficient to drive the inter-band transitions with an RF tone that is detuned from the inter-band coupling rate ($\Omega = \mu + \delta$). Doing so shifts the phase condition required to achieve maximum isolation, and the converted sideband shape becomes asymmetric. This effect is exacerbated when there is disorder in the mode hybridization. In this case, one inter-band transition better matches the RF drive frequency than the other, leading to more efficient conversion at this transition. The amplitudes of the converted sidebands differ, altering the power-matching condition required for isolation.

\begin{figure*}[hb]
    \centering
    \includegraphics[width=1\linewidth]{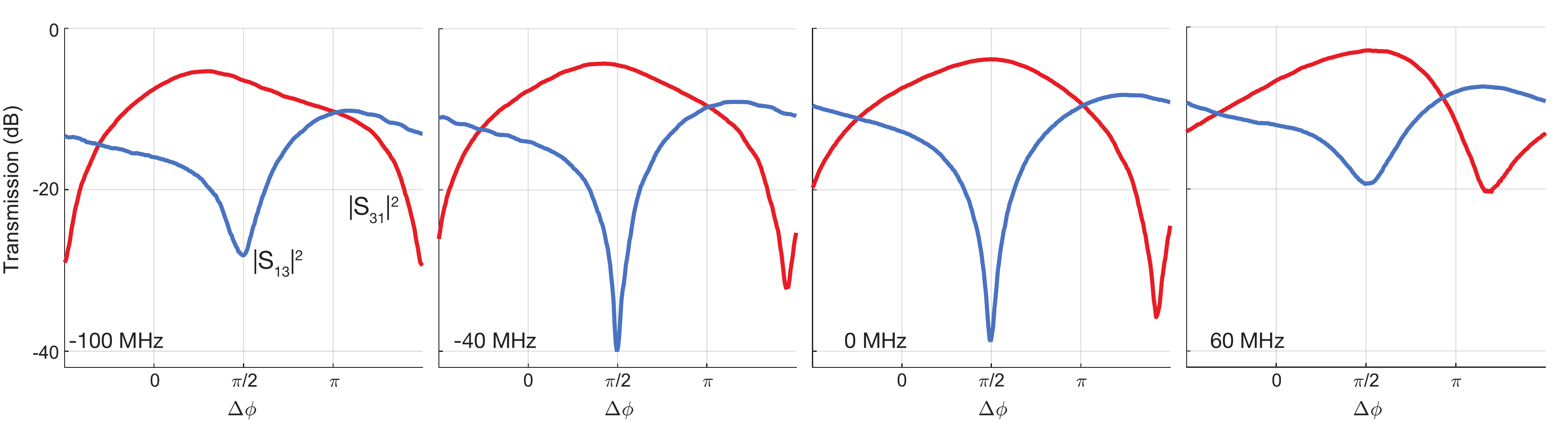}
    \caption{\textbf{Microwave Detuning Effects.} $|S_{31}|^2$ (red) and $|S_{13}|^2$ (blue) as a function of the phase condition, $\Delta\phi = 2\phi_1 - \phi_2$, for varied RF modulation detuning, as indicated on each panel.}
    \label{fig:Detuning_RF}
\end{figure*}

We detune the microwave drive from the ideal value and observe changes in lineshape and fluctuations in the isolation parameter, as depicted in Fig. \ref{fig:Detuning_RF}. These plots are measured for microwave drive amplitudes and frequencies as given in table \ref{tb:fitparams}.

\subsection{Laser Detuning and Isolation Bandwidth}
% \begin{figure*}[!t]
%     \centering
%     \includegraphics[width=0.6\linewidth]{Figures/LaserDetuning_2.pdf}
%     \caption{Isolation $I_{21}$ as a function of input signal detuning from super-mode frequency. We attribute the $5$ dB isolation ``floor'' to differences in optical signal transmission at far-blue detunings and consider this the zero-point of isolation. Therefore, $3$ dB bandwidth is approximated from an isolation of $8$ dB on this plot (at approximately $-100$ MHz detuning).}
%     \label{fig:Detuning_LSR}
% \end{figure*}
Detuning the optical signal from the resonance location has a similar effect as RF detuning on the $S_{ij}$ lineshapes and isolation parameters. 
% We hybridize the optical modes, predict the optical input frequencies using the fit technique described in section S1, and measure isolation as the input signal frequencies are shifted. Figure \ref{fig:Detuning_LSR} depicts the isolation parameter $S_{12}$ for this particular measurement ($P_1=$, $P_2=$, $\omega_1=$, $\omega_2=$). 
There is disorder in the hybridization, evidenced by misalignment between the maximum isolation and the resonance condition. From coupled mode theory, we can predict the isolation bandwidth at the ideal power matching condition $A_1$ and $A_2$, as given in the main text. For input and output frequencies detuned from the signal frequencies by $\Delta = \omega - \omega_i$, isolation parameters in the ideal system (\textit{i.e.,} no disorder in the mode hybridization) are given by:
\begin{equation}
    \begin{split}
        I_{31} &= 1 + \left(\frac{\kappa_1 + \kappa_3}{2\Delta}\right)^2 = 1 + \left(\frac{\gamma_2}{\Delta}\right)^2 \\
        I_{23,12} &= 1 + \left(\frac{\kappa_1 + 2\kappa_2 + \kappa_3}{4\Delta}\right)^2 = 1 + \left(\frac{\gamma_{1,3}}{\Delta}\right)^2
    \end{split}
\end{equation}
The 3 dB isolation bandwidth (FWHM) is given by twice the linewidth of the intermediary mode in the indirect transition pathway. In practice, our modes have disorder and exhibit some drift with RF drive power. We observed isolation that remained within a few dB of maximum over hundreds of MHz.
\clearpage

\section{Optical Power in the Cavity}
\begin{figure*}[!b]
    \centering
    \includegraphics[width=0.4\linewidth]{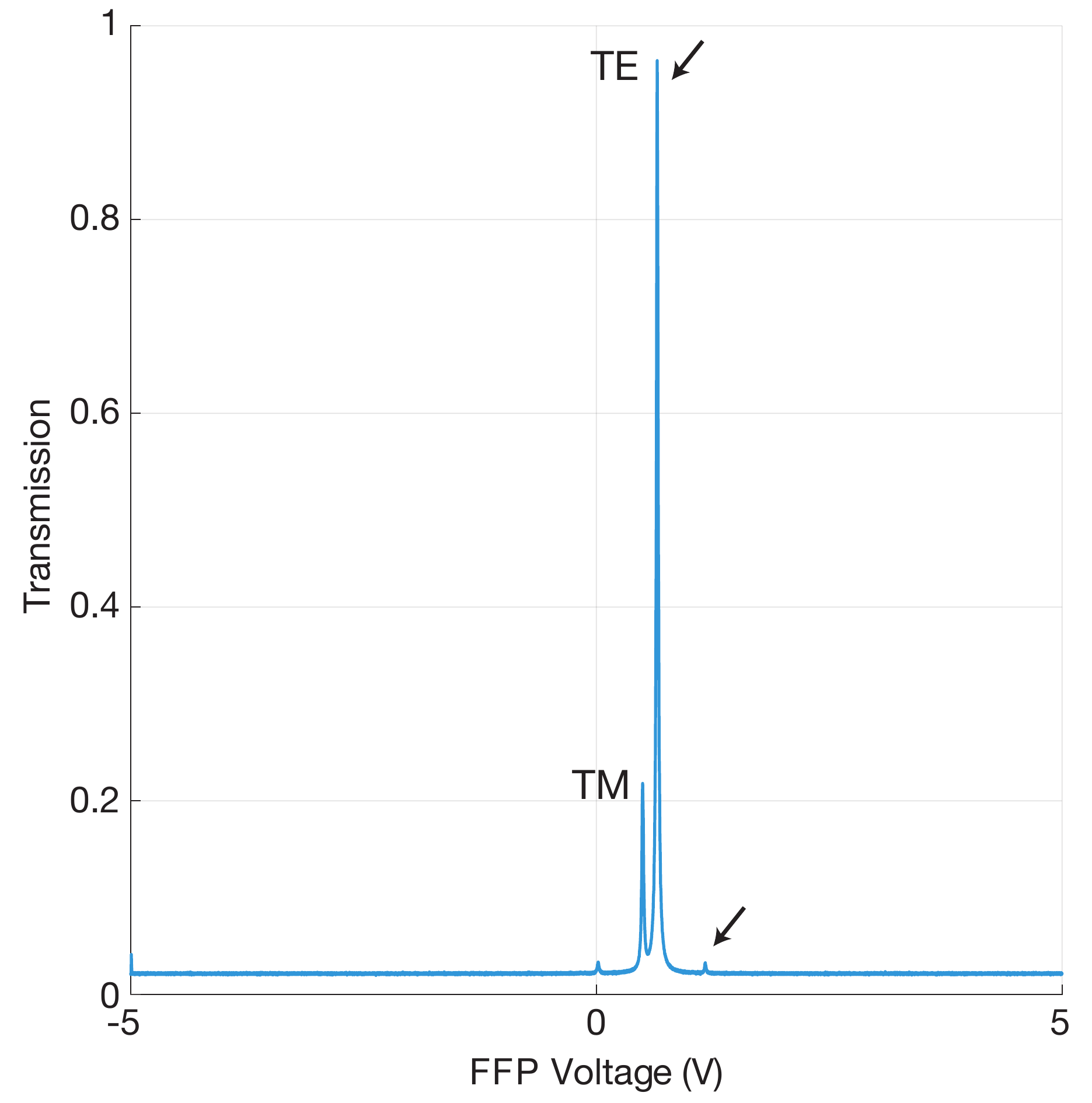}
    \caption{\textbf{Normalized EOM response filtered by Micron Optics Fiber Fabry-Perot (FFP).} By sweeping the voltage applied to the filter, the FFP cavity frequency is swept across the EOM carrier tone and sidebands. The central response exhibits two supported polarizations in the filter cavity, and the EOM carrier tone and first sidebands are visible (indicated by arrows). Here, modulation applied to the EOM is $1$ GHz.}
    \label{fig:FFP_EOMResponse}
\end{figure*}

We characterize optical power in the three resonators. Following the procedure set out by \cite{wentao2020}, we drive the EOM at various bias points, pass the output through a fiber Fabry-Perot filter (Micron Optics FFP), and record the transmission on a photodiode. The transmission is depicted in Fig.~\ref{fig:FFP_EOMResponse}. By comparing the sideband and carrier amplitudes after subtracting the detector noise floor, we identify how much light is scattered into sidebands for given modulation frequencies. We fit the scattering efficiency over a few drive frequencies to interpolate how much light is scattered into sidebands at the frequencies used in this experiment. By measuring the input power to the device ($2.246$ mW) and assuming equivalent coupling efficiency on the input and output of the device ($8.14\%$), we determine that approximately $5.1$ uW ($4.3$ uW and $3.6$ uW) of light propagates in the feed waveguide for signals at $\omega_1$ ($\omega_2$ and $\omega_3$).
We use the input-output formalism presented in section \ref{section:inputOutput} to determine the flux into the first racetrack coupled to the waveguide:

\begin{align}
    |\hat{a}_1|^2 &= \frac{\kappa_e}{\left|D_1+\frac{\mu_{12}^2}{D_2+\frac{\mu_{23}^2}{D_3}}\right|^2}|\alpha_{in}|^2\\
    E_{res} &= \hbar\omega|\hat{a}_1|^2\\
    D_i &= i\Delta' + \frac{\kappa_i}{2}
\end{align}
Substituting for the input flux, $|\alpha_{in}|^2$, in terms of optical power, we can solve for the circulating power in the first resonator:
\begin{equation}
\label{eq:P1}
    P_{1} = E_{res}\frac{v_g}{l} = \frac{\kappa_ev_g}{l\left|D_1+\frac{\mu_{12}^2}{D_2+\frac{\mu_{23}^2}{D_3}}\right|^2}P_{wg}
\end{equation}
In this expression $\kappa_1 = \kappa_e+\kappa_{1,i}$ and $P_{wg}$ refers to input signal power at a signal frequency $\omega_i$ propagating in the waveguide. We can similarly solve the input-output relations to obtain expressions for the power circulating in the second and third resonators:
\begin{equation}
\label{eq:P2}
    P_{2} = \left|\frac{-i\mu_{12}}{D_2+\frac{\mu_{23}^2}{D_3}}\right|^2P_1
\end{equation}
\begin{equation}
\label{eq:P3}
    P_3=\left|\frac{-i\mu_{23}}{D_3}\right|^2P_2
\end{equation}
From equations \ref{eq:P1},\ref{eq:P2},\ref{eq:P3}, using the average fit parameters obtained in section S1, we obtain power in resonators 1,2, and 3 for various signal frequencies (table \ref{tb:PowerInModes}).
\begin{table}[]
    \centering
    \caption{Average circulating power in resonators}
    \begin{tabular}{c|c|c|c|c}
    \hline
         Signal Frequency & Input Power ($\mu$W) & Res. 1 (mW) & Res. 2 (mW) & Res. 3 (mW) \\
    \hline\hline
         $\omega_1$ & 5.07 & 77 & 187 & 111 \\
         $\omega_2$ & 4.33 & 127 & 1.92 & 127 \\
         $\omega_3$ & 3.59 & 72 & 123 & 52\\
    \end{tabular}
    \label{tb:PowerInModes}
\end{table}
% \bibliography{papers_supplement}
\clearpage

\end{document}